\documentclass[aps,prx,twocolumn,superscriptaddress]{revtex4-1}
\usepackage{graphicx, float,hyperref}
\hypersetup{
     colorlinks   = true,
     allcolors    = [rgb]{0.0,0.0,1.0}
}
\begin{document}

\title{Pressure dependence of the structure and electronic properties of Sr$_3$Ir$_2$O$_7$}

\author{C. Donnerer}
\affiliation{London Centre for Nanotechnology, University College London, London, WC1H 0AH, UK}

\author{Z. Feng}
\affiliation{London Centre for Nanotechnology, University College London, London, WC1H 0AH, UK}

\author{J. G. Vale}
\affiliation{London Centre for Nanotechnology, University College London, London, WC1H 0AH, UK}
\affiliation{Laboratory for Quantum Magnetism, Ecole Polytechnique Federal de Lausanne (EPFL), 1015 Lausanne, Switzerland}

\author{S.N. Andreev}
\affiliation{Theoretical Physics and Applied Mathematics Department, Ural Federal University, 620002 Ekaterinburg, Russia}

\author{I.V. Solovyev}
\affiliation{Theoretical Physics and Applied Mathematics Department, Ural Federal University, 620002 Ekaterinburg, Russia}
\affiliation{Computational Materials Science Unit, National Institute for Materials Science, 1-1 Namiki, Tsukuba, Ibaraki 305-0044, Japan}

\author{E. C. Hunter}
\affiliation{SUPA, School of Physics and Astronomy, and Centre for Science at Extreme Conditions, The University of Edinburgh, Mayfield Road, Edinburgh EH9 3JZ, United Kingdom}

\author{M. Hanfland}
\affiliation{European Synchrotron Radiation Facility, CS 40220, 38043 Grenoble Cedex 9, France}

\author{R. S. Perry}
\affiliation{London Centre for Nanotechnology, University College London, London, WC1H 0AH, UK}

\author{H. M. R\o{}nnow}
\affiliation{Laboratory for Quantum Magnetism, Ecole Polytechnique Federal de Lausanne (EPFL), 1015 Lausanne, Switzerland}

\author{M. I. McMahon}
\affiliation{SUPA, School of Physics and Astronomy, and Centre for Science at Extreme Conditions, The University of Edinburgh, Mayfield Road, Edinburgh EH9 3JZ, United Kingdom}
\affiliation{Research Complex at Harwell, Didcot, Oxon OX11 0FA, United Kingdom}

\author{V. V. Mazurenko}
\affiliation{Theoretical Physics and Applied Mathematics Department, Ural Federal University, 620002 Ekaterinburg, Russia}

\author{D. F. McMorrow}
\affiliation{London Centre for Nanotechnology, University College London, London, WC1H 0AH, UK}

\begin{abstract}
We study the structural evolution of Sr$_3$Ir$_2$O$_7$ as a function of pressure using x-ray diffraction. At a pressure of 54 GPa at room temperature, we observe a first-order structural phase transition, associated with a change from tetragonal to monoclinic symmetry, and accompanied by a 4\% volume collapse. Rietveld refinement of the high-pressure phase reveals a novel modification of the Ruddlesden-Popper structure, which adopts an altered stacking sequence of the perovskite bilayers. As the positions of the oxygen atoms could not be reliably refined from the data, we use density functional theory  (local-density approximation+$U$+spin orbit) to optimize the crystal structure, and to elucidate the electronic and magnetic properties of Sr$_3$Ir$_2$O$_7$ at high pressure. In the low-pressure tetragonal phase, we find that the in-plane rotation of the IrO$_6$ octahedra increases with pressure. The calculations further indicate that a bandwidth-driven insulator-metal transition occurs at $\sim$20 GPa, along with a quenching of the magnetic moment. In the high-pressure monoclinic phase, structural optimization resulted in complex tilting and rotation of the oxygen octahedra, and strongly overlapping $t_{2g}$ and $e_g$ bands. The $t_{2g}$ bandwidth renders both the spin-orbit coupling and electronic correlations ineffectual in opening an electronic gap, resulting in a robust metallic state for the high-pressure phase of Sr$_3$Ir$_2$O$_7$.
\end{abstract}

\maketitle

\section{Introduction} 

Understanding the metal-insulator transition (MIT) in transition metal oxides is pivotal in explaining the rich phenomenology that these compounds display \cite{imada1998metal}. Recently, strong spin-orbit coupling has been shown to lead to novel MITs and associated electronic and magnetic states in oxides containing a $5d$ element \cite{krempa2014correlated}. While the strongly correlated $3d$ oxides are described as Mott insulators, a range of mechanisms have been proposed to account for the insulating ground states of the $5d$ oxides \cite{krempa2014correlated, kim2008novel, moon2008dimensionality, arita2012ab}. In particular, it remains unclear how the metal-insulator transition evolves in the correlated, strong spin-orbit coupling regime relevant to these materials. 

In the Ruddlesden-Popper series of perovskite iridates Sr$_{n+1}$Ir$_n$O$_{3n+1}$ (where $n$ is the number of Ir-O layers, separated by Sr-O layers), the large octahedral crystal-field splitting and spin-orbit coupling produce a $J_{\text{eff}}=1/2$ band for Ir$^{4+}$ ($5d^5$). For Sr$_2$IrO$_4$ ($n=1$), it was proposed that the narrow $J_{\text{eff}}=1/2$ band is split by the modest electronic correlations to create a spin-orbit Mott insulator \cite{kim2008novel, kim2009phase, jackeli2009mott}. The ideal realization of the $J_{\text{eff}}=1/2$ state carries a magnetic moment of 1$\mu_B$, however, experimental values of 0.5$\mu_B$ \cite{cao2002anomalous, dhital2012spin} suggest significant degrees of itineracy. The perovskite iridates are therefore located in the intermediate-coupling regime, and the insulating state has been proposed to exhibit both Slater and Mott characteristics \cite{arita2012ab, hsieh2012observation, watanabe2014theoretical}. The bilayer compound Sr$_3$Ir$_2$O$_7$ ($n=2$) marginally retains the insulating $J_{\text{eff}}=1/2$ state \cite{kim2012dimensionality, boseggia2012antiferromagnetic, zhang2013effective}. The effective change in dimensionality in the bilayer structure places Sr$_3$Ir$_2$O$_7$ in close proximity to a metal-insulator transition, as SrIrO$_3$ ($n=\infty$) is found to be metallic \cite{moon2008dimensionality}.

The charge gap of $\sim$130 meV in Sr$_3$Ir$_2$O$_7$ \cite{okada2013imaging} can be closed by introducing mobile carriers through substituting La$^{3+}$ on Sr$^{2+}$ sites \cite{li2013tuning, de2014coherent, he2015spectroscopic}. This weakens  electron 
correlations such that a conventional Fermi liquid state is recovered. In this context, application of pressure is an attractive control parameter, as it allows to continuously and isoelectronically tune the salient energy scales. Zhao \emph{et al.} reported a second-order structural transition in Sr$_3$Ir$_2$O$_7$ at 14 GPa, accompanied by a reduction in resistivity \cite{li2013tuning, zhao2014pressure}. However, high-pressure transport measurements by Zocco \emph{et al.} \cite{zocco2014persistent} have seen no evidence for this transition, and instead report a gradual reduction of the electronic gap up to 104 GPa. While the authors do not report definitive evidence for a transition to a
metallic state, they concede that pressure gradients could affect their data and that an insulator-metal transition may occur at lower pressures.

Here, we present a comprehensive study of the pressure dependence of the structure of Sr$_3$Ir$_2$O$_7$ up to 61 GPa, using powder x-ray diffraction (XRD) at ambient temperature and at 20 K. We identify a first-order structural transition to a monoclinic symmetry accompanied by a volume collapse of 4\% at 54 GPa at room temperature. Rietveld refinement suggests that the stacking sequence of the two perovskite bilayers is altered in the high-pressure phase, resulting in a novel modification of the Ruddlesden-Popper structure. In contrast to the study by Zhao \emph{et al.} \cite{zhao2014pressure}, no evidence for a second-order structural transition around 14 GPa was found either at room temperature or at 20 K.

As the oxygen atoms scatter only weakly compared to the electron-heavy Sr and Ir atoms, a reliable refinement of the oxygen positions was not possible from the diffraction data. We employed density functional theory [local-density approximation (LDA)+$U$+spin-orbit (SO)] calculations to refine the structure of Sr$_3$Ir$_2$O$_7$ at high pressure and to examine the evolution of the electronic and magnetic states. The calculations suggest that the in-plane rotation of IrO$_6$ octahedra increases with pressure in the tetragonal phase. The decrease of the Ir-O bond distance, however, widens the $t_{2g}$ bands, such that an insulator-metal transition occurs at $\sim$20 GPa. For the monoclinic high-pressure phase, optimization of the oxygen positions leads to complex rotations and tilting of octahedra, which considerably lowers the total energy with respect to the structure proposed by experiment. In this high-pressure structure, we find strongly overlapping $t_{2g}$ and $e_g$ bands. Due to the wide $t_{2g}$ bands, both the spin-orbit coupling and electronic correlations become ineffectual in opening an electronic gap and we obtain a metallic state for the high-pressure phase.

\section{High-Pressure x-ray diffraction}

\subsection{Experimental method}

We conducted high-pressure x-ray diffraction experiments at beamlines ID09a (European Synchrotron Radiation Facility) and I15 (Diamond Light Source). Single crystals of Sr$_3$Ir$_2$O$_7$ were flux grown as described in Ref. \onlinecite{boseggia2012antiferromagnetic} and characterized by SQUID magnetometry. We used symmetrical diamond-anvil cells, fitted with Boehler-Almax diamonds with a culet diameter of 250 $\mu$m. The gasket material was rhenium, indented to 30 $\mu$m thickness, into which a 100-$\mu$m-diameter hole was laser-drilled. The Sr$_3$Ir$_2$O$_7$ single crystals were ground into powders and loaded with Helium as the pressure transmitting medium. Pressure was measured \emph{in situ} using the SrB$_4$O$_7$:Sm$^{2+}$ fluorescence pressure scale \cite{datchi1997improved}. The diffraction patterns were recorded on MAR345 and MAR555 detectors and integrated using the Fit2D software \cite{Fit2D}. Full profile Rietveld refinements were performed using the FullProf software \cite{rodriguez1990fullprof}.

\begin{figure}[!t]
\centering \includegraphics[width=\linewidth]{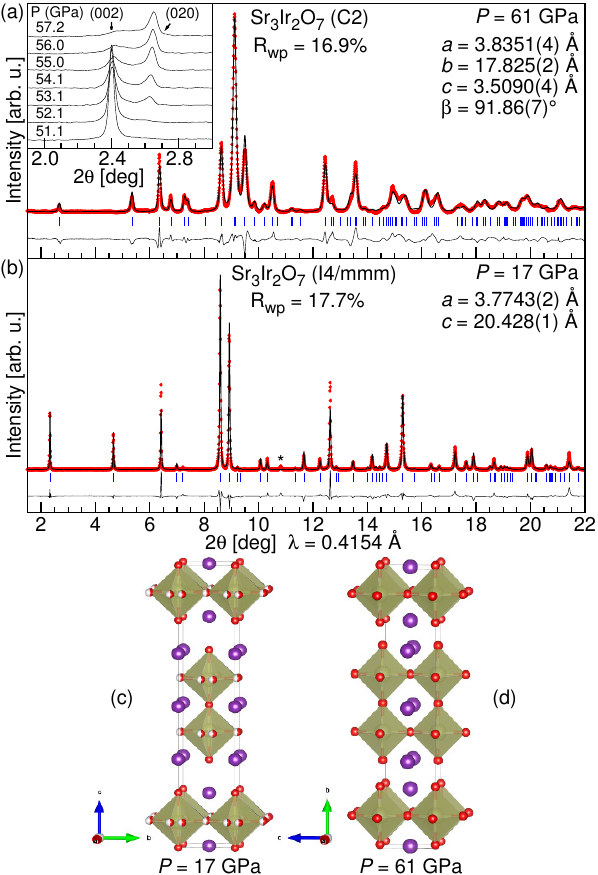}
\caption{(Color online). Integrated XRD spectra of Sr$_3$Ir$_2$O$_7$ at high pressure at ambient temperature. (a) Sr$_3$Ir$_2$O$_7$ in its monoclinic phase (space group $C2$) at 61 GPa; (d) the corresponding structure. (b) Tetragonal phase of Sr$_3$Ir$_2$O$_7$ (space group $I4/mmm$) at 17 GPa; (c) the corresponding structure. In (a) and (b) the red circles are background-subtracted data points, the black solid line presents the Rietveld refinement, blue tick marks are the calculated Bragg reflections, and the residuals of the refinement are shown by the black line beneath. Reflections from SrB$_4$O$_7$:Sm$^{2+}$ are marked by an asterisk. The inset in (a) shows the pressure evolution of small-angle reflections. In (c) and (d) the iridium atoms are shown in bronze, the oxygen atoms in red and the strontium atoms in purple. The structures in (c) and (d) were visualised via VESTA \cite{VESTA}.}
\label{XRD}
\end{figure}

\subsection{Results and discussion}

Diffraction patterns of Sr$_3$Ir$_2$O$_7$ at high pressure are shown in Fig. \ref{XRD}. At ambient temperature in a helium pressure-transmitting medium, Sr$_3$Ir$_2$O$_7$ remained in its tetragonal phase (space group $I4/mmm$) up to pressures of 53 GPa. It has been argued that the rotations of the oxygen octahedra are correlated, resulting in a 90$^{\circ}$ twinned structure of pseudo-orthorhombic space groups $Bbcb$ and $Acaa$ \cite{matsuhata2004crystal}. As our measurements were performed on powder samples, the tetragonal $I4/mmm$ structural model effectively averages over the two space groups. Rietveld refinements of our data confirm the $I4/mmm$ structural model to be adequate. Crucially, no second-order structural transition was observed near 14 GPa. We used helium as the pressure-transmitting medium (at room temperature and 20 K), and not neon as in Ref. \onlinecite{zhao2014pressure} (at 25 K). The discrepancy between our studies may  therefore arise from nonhydrostatic pressure conditions in the previous study. In our experiments, the pressure was determined by SrB$_4$O$_7$:Sm$^{2+}$ fluorescence. No significant broadening of the SrB$_4$O$_7$:Sm$^{2+}$ fluorescence line shapes was observed, indicating that hydrostatic pressure conditions were maintained. At pressures above 54 GPa, new reflections started to emerge. Further increase in pressure resulted in the complete disappearance of the $I4/mmm$ reflections. The new reflections were indexed in a $C$-centered monoclinic system, space group $C2/m$, $C2$, or $Cm$. Indexing the reflections in a higher symmetry proved not to be possible. This high-pressure phase remained stable up to 61 GPa, the maximum pressure reached in this study. Upon decompression, the monoclinic signature disappeared below 49 GPa, and Sr$_3$Ir$_2$O$_7$ reverted to its tetragonal phase.

\begin{figure}[htb]
\includegraphics[width=\linewidth]{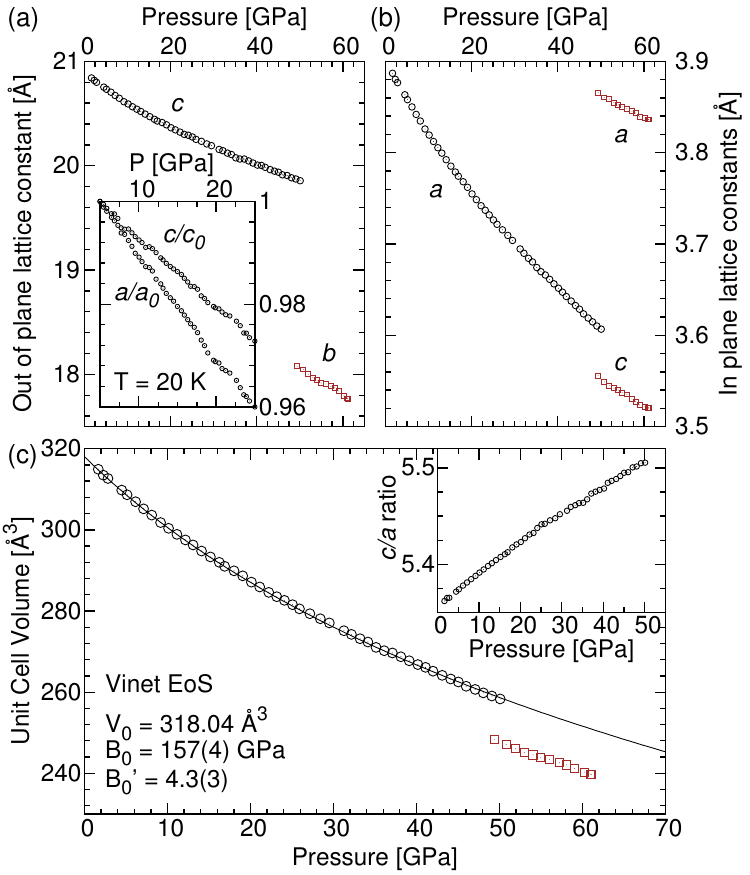}
\caption{(Color online). Equation of state of Sr$_3$Ir$_2$O$_7$.  (a) and (b) Lattice parameters as a function of pressure at 298 K. The inset in (a) shows the normalized lattice constants as a function of pressure at 20 K. (c) Equation of state of Sr$_3$Ir$_2$O$_7$ at 298 K. The solid line is a Vinet-type fit to the observed equation of state of Sr$_3$Ir$_2$O$_7$. The inset in (c) shows the $c/a$ ratio as a function of pressure at 298 K.}
\label{EoS}
\end{figure}

The high-pressure phase of Sr$_3$Ir$_2$O$_7$ departs strongly from the Ruddlesden-Popper structure. Most notably, the reflection with the largest $d$ spacing, (002) at $2\theta = 2.4^\circ$, disappears [see inset Fig. \ref{XRD}(a)]. All new reflections possess smaller $d$ spacings, indicating that the long $c$ axis of tetragonal Sr$_3$Ir$_2$O$_7$ is not retained. Figure \ref{EoS} shows the refined lattice parameters as a function of pressure. In the tetragonal phase, the compression of Sr$_3$Ir$_2$O$_7$ is anisotropic; the $c$ axis exhibits a lower compressibility than the $a$=$b$ axes. The $c/a$ ratio thus increases with pressure by 3\% to a critical value, where the system undergoes a structural phase transition. We indexed the new phase in a monoclinic symmetry, spacegroup $C2/m$, $C2$, or $Cm$, with lattice parameters $a=3.8351(4)$ \AA, $b=17.825(2)$ \AA, $c=3.5090(4)$ \AA, and $\beta=91.86(7)^{\circ}$ at 61 GPa. This corresponds to a volume collapse of 4\% at the phase transition. A discontinuity of 2 \AA{} in the long $c$ axis of the unit cell (labeled the $b$ axis in the high-pressure phase) occurs at the structural transition. The lower compressibility along the $c$ axis in the tetragonal phase appears to be alleviated by the phase transition to monoclinic symmetry. A Vinet equation of state was fitted to the pressure-volume data of the tetragonal phase, yielding a bulk modulus of $B_0 =157(4)$ GPa [$B_0' = 4.3(3)$], in agreement with the literature \cite{zhao2014pressure}.

To understand the transition at an atomic level, we attempted structural refinement of the high-pressure phase through Rietveld analysis. In general, the calculated intensities reproduce adequately the experimental intensities when one of the perovskite bilayers of Sr$_{3}$Ir$_{2}$O$_{7}$ is translated in plane, by half a unit cell [see Fig. \ref{XRD}(d)]. This may explain the discontinuity in the lattice parameter of the long axis of the unit cell. While in its tetragonal phase the compressibility along the $c$ axis appears to be limited, this is alleviated in the monoclinic phase by an alteration in the stacking sequence of the perovskite bilayers. Furthermore, the Sr$^{2+}$ atoms are arranged within a single plane, parallel to the $a$ axis in the unit cell. To accommodate this configuration, the lattice parameter $a$ increases, such that the in-plane lattice constants ($a,c$) of the monoclinic structure differ by 10\%, and the angle spanned between $a$ and $c$ is 91.8$^\circ$. 

This high-pressure structure could be described in the space group $C2/m$. The presence of a mirror plane perpendicular to the $b$ axis minimizes the number of free parameters to two $y$ coordinates for Ir and Sr (the second Sr atom is fixed at $y=0.5$). However, the fit of the calculated diffraction pattern was deemed unsatisfactory. We therefore decided to remove the mirror plane and adopt $C2$ symmetry. This allowed to move atoms independently along the $y$ axis, through two adjustable $y$ coordinates for both Ir and Sr atoms (here, the third Sr atom remained at $y=0.5$). This provided the best fit to the data. The weighted profile Rietveld $R$ factor $R_{wp}$ of the monoclinic structure at 61 GPa then became comparable to the one obtained from the Rietveld refinement of the tetragonal phase at 17 GPa. Thus our proposed structural model for the high-pressure phase provides as good a description of the data as the $I4/mmm$ structure at lower pressures. We note that the visually worse fit of the calculated intensities at 61 GPa at least partly originates from the difficulty of accurately describing the background at very high pressures.

As the oxygen atoms scatter only weakly compared to the electron-heavy Ir and Sr atoms, a precise refinement of the oxygen positions was not possible for the high-pressure phase. Small deviations from the proposed positions [as shown in Fig. \ref{XRD}(d)] through rotations or tilting cannot be excluded or confirmed within the quality of our data.

\section{Results of the first-principles calculations}

\subsection{Numerical methods}

To investigate the pressure dependence of the electronic and magnetic properties of Sr$_{3}$Ir$_{2}$O$_{7}$ we performed first-principles calculations with two complementary approaches: the full-potential linearized-augmented-plane wave (FP-LAPW) method, as realized in the Elk package \cite{singh2006planewaves}; and the plane wave pseudopotential method, as implemented in the QUANTUM-ESPRESSO simulation package \cite{giannozzi2009quantum}. In these calculations we used the LDA, and its popular refinement taking into account the on-site Coulomb interaction and spin-orbit coupling (LDA+$U$+SO) \cite{solovyev1998hund}. In pseudopotential calculations, we set an energy cutoff of 400 eV in the plane-wave basis construction and the energy convergence criteria of 10$^{-4}$ eV. For the Brillouin-zone integration, a 6 $\times$ 6 $\times$ 6 Monkhorst-Pack mesh was used.

\begin{figure}[!h]
\includegraphics[width=0.5\textwidth,angle=0]{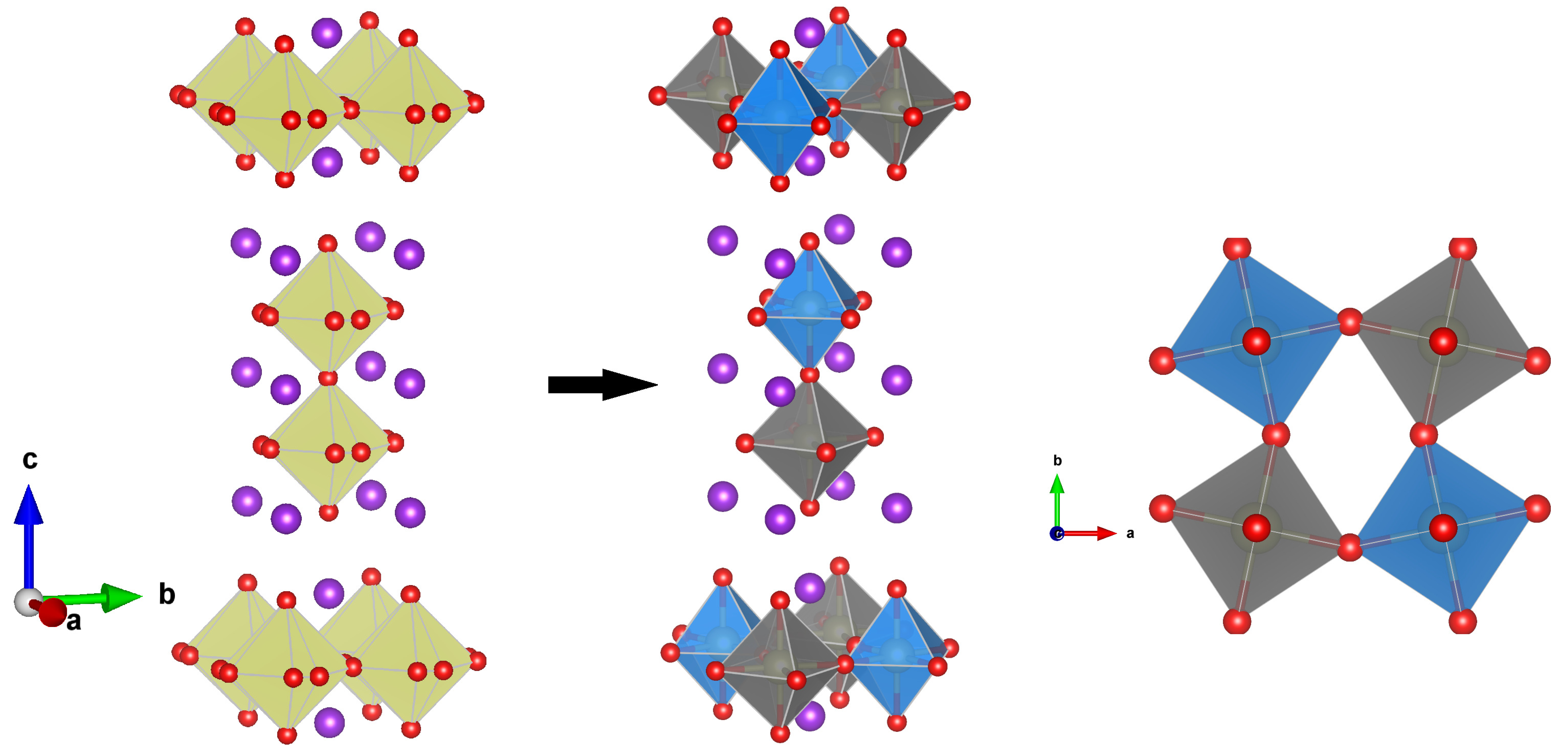}
\caption{(Color online). (Left) Structure of Sr$_3$Ir$_2$O$_7$ in the ambient, tetragonal phase (space group $I4/mmm$) with oxygen disorder. (Middle) $Bbcb$ structure of Sr$_3$Ir$_2$O$_7$ used in first-principles calculations. The difference between blue and gray IrO$_6$ octahedra is the octahedral rotation around the $z$ axis. The red and purple spheres denote oxygen and strontium atoms, respectively. (Right) $xy$-plane projection of the $Bbcb$ structure. Crystallographic plots were made with the VESTA software \cite{VESTA}.}
\label{structuresBbcb}
\end{figure}  

\subsection{Structural and electronic properties at ambient pressure} 

We first discuss the LDA and LDA+SO results for the ambient pressure phase of Sr$_3$Ir$_2$O$_7$. In these calculations, we started from the $I4/mmm$ crystal structure with oxygen disorder \cite{subramanian1994single} and then doubled the planar basis lattice vectors and deleted specific planar oxygen atoms in the supercell, to achieve the $Bbcb$ space group (Fig. \ref{structuresBbcb}).

\begin{figure}[!t]
\includegraphics[width=0.45\textwidth,angle=0]{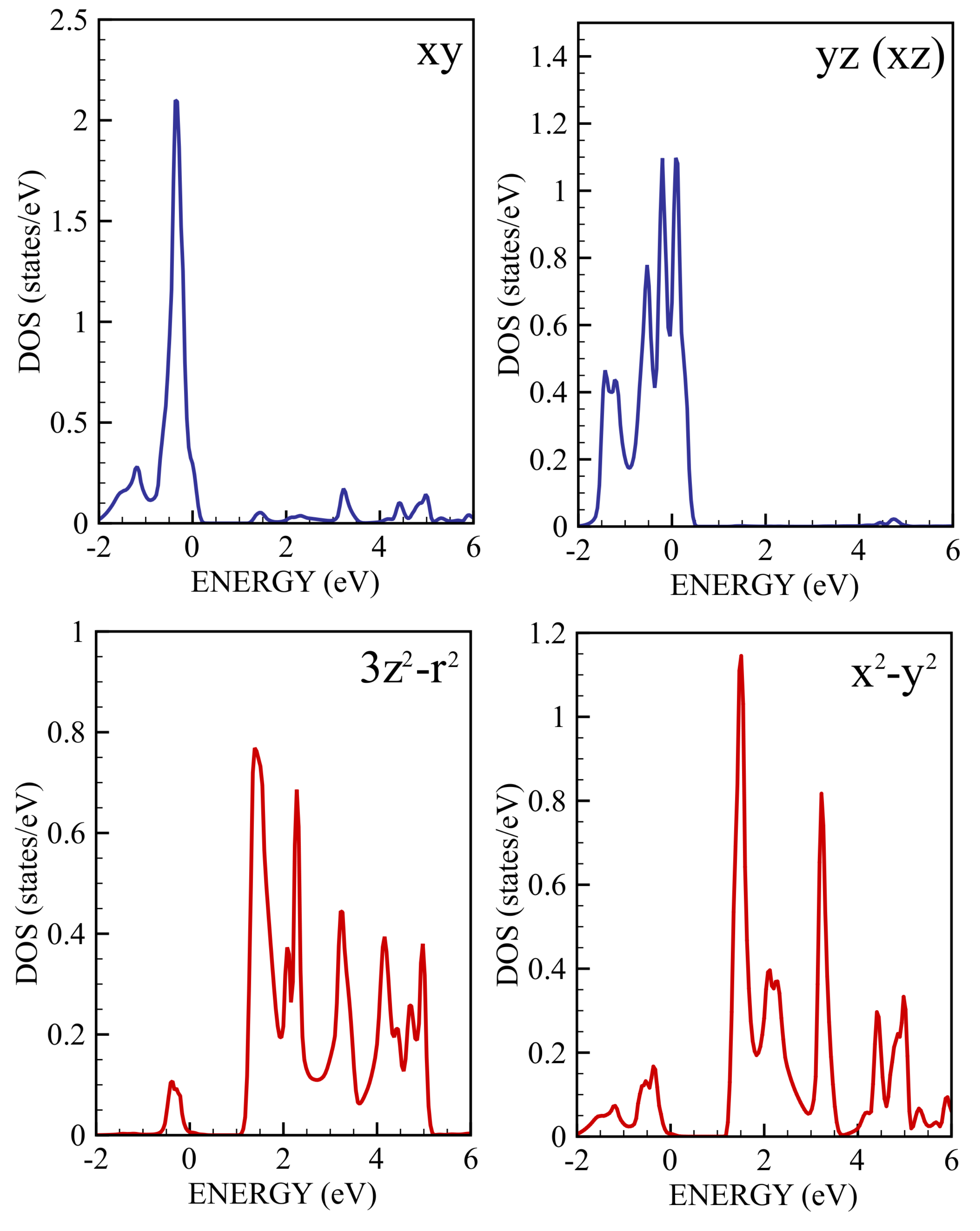}
\caption{(Color online). The partial densities of states calculated for Sr$_3$Ir$_2$O$_7$ in the $Bbcb$ structure at ambient pressure using the LDA method.}
\label{pdosLDA_ap}
\end{figure} 

From the calculated partial densities of states presented in Fig. \ref{pdosLDA_ap}, the splitting between $t_{2g}$ and $e_{g}$ states can be estimated to be about 3.3 eV, in good agreement with the experimental value of 3.5 eV \cite{moretti2014crystal}. We find that the $5d$ Ir states of $xy$ character are filled and the $yz$ ($xz$) orbitals are partially occupied. The $e_{g}$ band is fully located in the unoccupied part of the electronic spectrum due to much stronger hybridization with oxygen states. The obtained level picture agrees with symmetry analysis of the Sr$_3$Ir$_2$O$_7$ system \cite{cao2002anomalous}. As we show below, the degeneracy of $xz$ and $yz$ states plays an important role for the orbital magnetism. 

\begin{figure*}[]
\includegraphics[width=0.75\textwidth,angle=0]{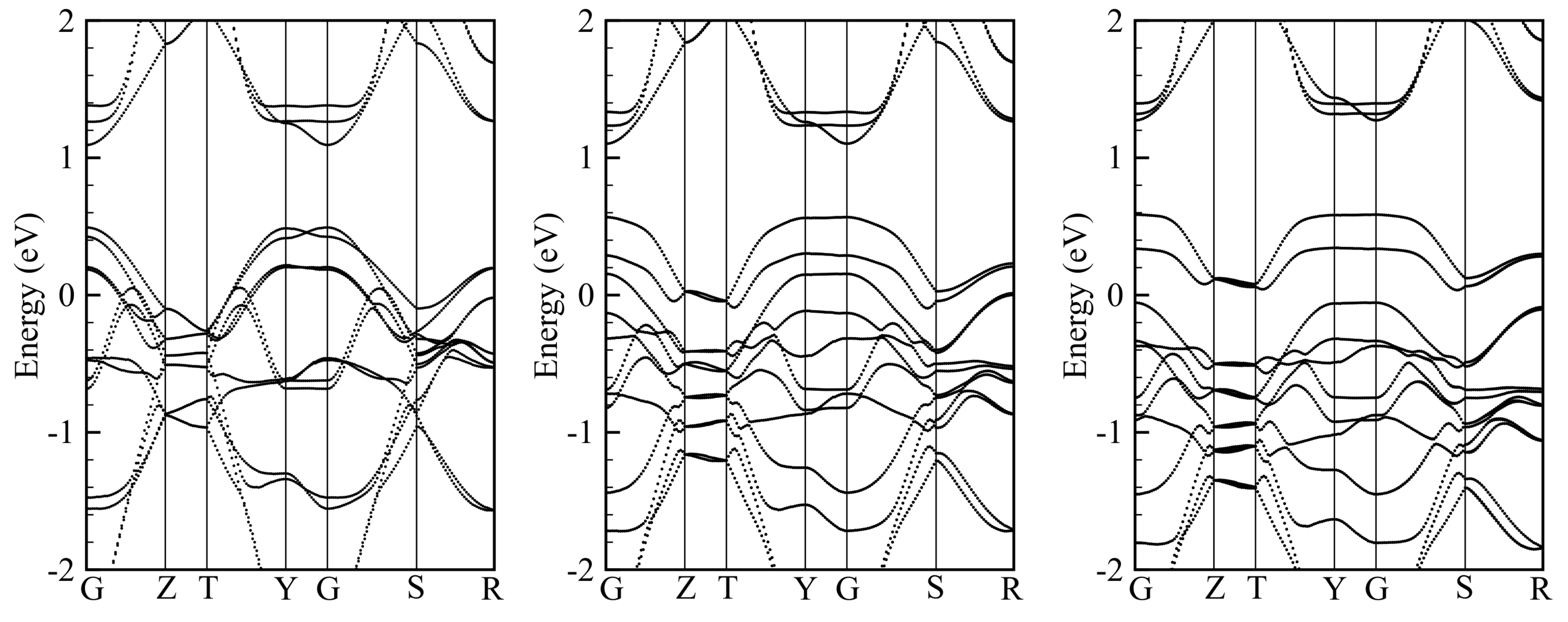}
\caption{(Color online). The band structures of Sr$_3$Ir$_2$O$_7$ at ambient pressure obtained from LDA (left), LDA+SO (middle) and LDA+$U$+SO (right) calculations using the $Bbcb$ structure. The coordinates of the symmetry points, G=( 0, 0, 0), Z=( 0, $\frac{1}{2}$, 0), T=( 0, $\frac{1}{2}$, -$\frac{a}{2c}$ ), Y=( 0, 0, -$\frac{a}{2c}$), S=( $\frac{1}{2}$, 0, $\frac{a}{4c}$), R=( $\frac{1}{2}$, $\frac{1}{2}$, -$\frac{a}{4c}$) are in units of $\frac{2\pi}{a}$, where $a$ = 5.5 \AA \, and $c$ = 20.87 \AA.}
\label{bandLDA_ap}
\end{figure*} 

The comparison of the LDA and LDA+SO band structures (Fig. \ref{bandLDA_ap}, left) shows that the spin-orbit coupling is responsible for the formation of four doubly degenerate nearly separated narrow bands at the Fermi level. The corresponding bandwidth can be estimated as 0.5 eV. Given the large crystal field, an {\it effective} total angular momentum approach can be used to construct a minimal microscopic model of the insulating ground state of Sr$_3$Ir$_2$O$_7$, as proposed in Ref. \onlinecite{kim2008novel}. However, in contrast to the ideal $J_{\text{eff}}=1/2$ state described in Ref. \onlinecite{kim2008novel}, we have a more complicated situation. The splitting of the $t_{2g}$ manifold results in a mixing of $J_{\text{eff}}=1/2$ and $J_{\text{eff}}=3/2$ bands, which should be taken into account within a microscopic analysis \cite{moretti2014orbital}.

We employed the LDA+$U$+SO method to first reproduce the insulating ground state and magnetic properties of Sr$_{3}$Ir$_{2}$O$_{7}$ at ambient pressure. The obtained band structure is illustrated in Fig. \ref{bandLDA_ap}. In our simulations, the on-site Coulomb interaction $U$ and intra-atomic exchange parameter $J_{H}$ were chosen to be 1.8 and 0.3 eV, respectively, as discussed in Ref. \onlinecite{moser2014electronic}. The electronic gap obtained from the calculations is 110 meV, in reasonable agreement with the experimental estimate of 130 meV \cite{okada2013imaging}. The calculated spin and orbital magnetic moments are M$_{S} = 0.14\mu_{B}$ and M$_{L} = 0.24\mu_{B}$, respectively, in good agreement with the total magnetic moment of 0.52(8)$\mu_{B}$ measured by neutron diffraction \cite{dhital2012spin}. It is interesting to note that the calculated orbital moment of Sr$_{3}$Ir$_{2}$O$_{7}$ is comparable to the one calculated for Sr$_2$IrO$_4$ \cite{solovyev2015validity}, while the spin moment  is two times larger.

We have simulated different $c$-oriented magnetic configurations of Sr$_3$Ir$_2$O$_7$: ferromagnetic, and antiferromagnetic of $G$, $C$, and $A$ types. We find that the $c$-axis antiferromagnetic configuration of $G$-type minimizes the total energy of the system, in agreement with experiments \cite{kim2012dimensionality, boseggia2012antiferromagnetic}. The calculations for other types of the magnetic order revealed strong changes in electronic and magnetic properties of Sr$_3$Ir$_2$O$_7$. In general, the magnetic moments and energy gap values are suppressed in comparison with those obtained for $G$-type structure. 

In turn, the in-plane antiferromagnetic solution with weak ferromagnetism also corresponds to the excited states of the Sr$_3$Ir$_2$O$_7$ system. The corresponding energy difference between in-plane and out-of-plane $G$-type solutions is 353 meV (per unit cell containing four Ir atoms). The calculated orbital and spin magnetic moments of Ir atom in the in-plane magnetic configuration are 0.186$\mu_{B}$ and 0.063$\mu_{B}$, respectively. The canting angles of the orbital and spin moments are equal to 15.2$^{\circ}$ and 18.4$^{\circ}$, respectively. As in the case of Sr$_2$IrO$_4$ \cite{solovyev2015validity} we found a strong contribution from the oxygen atoms to the resulting net magnetic moment.

\subsection{Structural and electronic properties at high pressure} 

To understand the evolution of the electronic and magnetic properties of Sr$_3$Ir$_2$O$_7$ as a function of pressure, we have performed LDA+$U$+SO calculations for the $Bbcb$ structure, using the experimental lattice parameters (Fig. \ref{EoS}). For each pressure point, the crystal structure was optimized. The resulting structural, electronic, and magnetic properties of Sr$_3$Ir$_2$O$_7$ as a function of pressure are shown in Table \ref{Pressuredata}.

Our calculations reveal that the compression of the unit cell is accompanied by an increase of the in-plane rotation angle of the IrO$_6$ octahedra, while the Ir-O distance decreases. Intuitively, this can be understood from the experimentally observed anisotropic compression of Sr$_3$Ir$_2$O$_7$, where the compressibility along the $a$ and $b$ axes is higher than that along the $c$ axis. At ambient pressure, the rotation angle is about 11.8$^{\circ}$, while the high-pressure solution at 50 GPa reveals an octahedral rotation of 16.3$^{\circ}$. 

\begin{figure}[!b]
\includegraphics[width=\linewidth]{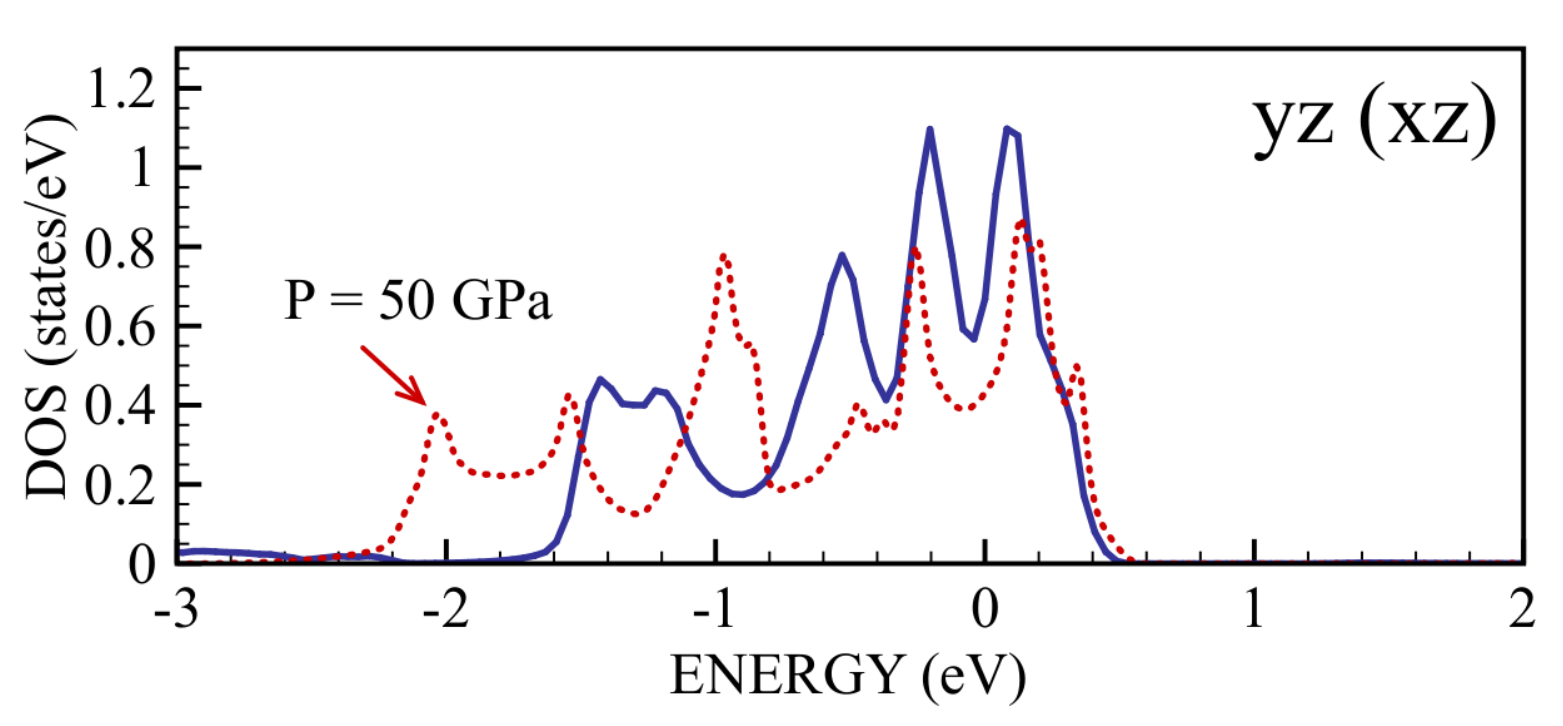}
\caption{(Color online). Comparison of the partial densities of states of Sr$_3$Ir$_2$O$_7$ in the $Bbcb$ structure, calculated using the LDA method at ambient pressure (blue solid line) and 50 GPa (red dotted line).}
\label{dos_amb_high}
\end{figure}

For Sr$_2$IrO$_{4}$ and Ba$_{2}$IrO$_{4}$, it was demonstrated that the in-plane rotation of the IrO$_6$ octahedra with fixed Ir-O distance leads to a more localized electronic structure \cite{solovyev2015validity}. Therefore the width of the $t_{2g}$ band was found to be narrower for Sr$_{2}$IrO$_4$ than for Ba$_2$IrO$_4$. For Sr$_3$Ir$_2$O$_7$, applying pressure also decreases the Ir-O distance, which leads to a strong increase of the LDA bandwidth for $t_{2g}$ states. Figure \ref{dos_amb_high} shows that the spectral function for degenerate $xz$ and $yz$ states retains its structure at high pressure; however, the width of the spectral function notably increases at 50 GPa.

\begin{table}[!h]
\centering
\caption{Pressure dependence of the structural, electronic, and magnetic properties of Sr$_3$Ir$_2$O$_7$, using the $Bbcb$ structure. The results were obtained with LDA+$U$+SO calculations, as described in the text. $\alpha$ is the angle of the in-plane Ir-O-Ir bond. E$_{gap}$ is the value of the energy gap. The lattice parameters corresponding to the pressure of 60 GPa were obtained by extrapolation of the experimental data.}
\label{Pressuredata}
\begin {tabular}{ccccccc}
\hline
P (GPa) & $\alpha$ & d$_{Ir-O}$ (\AA) & E$_{gap}$ (meV) & M$_{S}$ ($\mu_B$) & M$_{L}$ ($\mu_B$)  \\
  \hline
  0    & 156.4 & 1.99  & 110 & 0.14  & 0.24  \\
  10 &   155.4  &  1.95   &  62  & 0.14  &   0.21 \\
  20 & 153.1 & 1.93  & -     & 0.11 & 0.14   \\
  30  & 150.9 & 1.91  & -    &  0.07  &  0.08  \\
  40  & 148.4 & 1.89  & -    &   0.05   & 0.05  \\
  50 & 147.4 & 1.88  &  -    & 0.04 & 0.04   \\
  60 & 144.6 & 1.87  & -     & 0.04 & 0.03 \\
  \hline
\end {tabular}
\end {table}

\begin{figure}[!t]
\includegraphics[width=\linewidth]{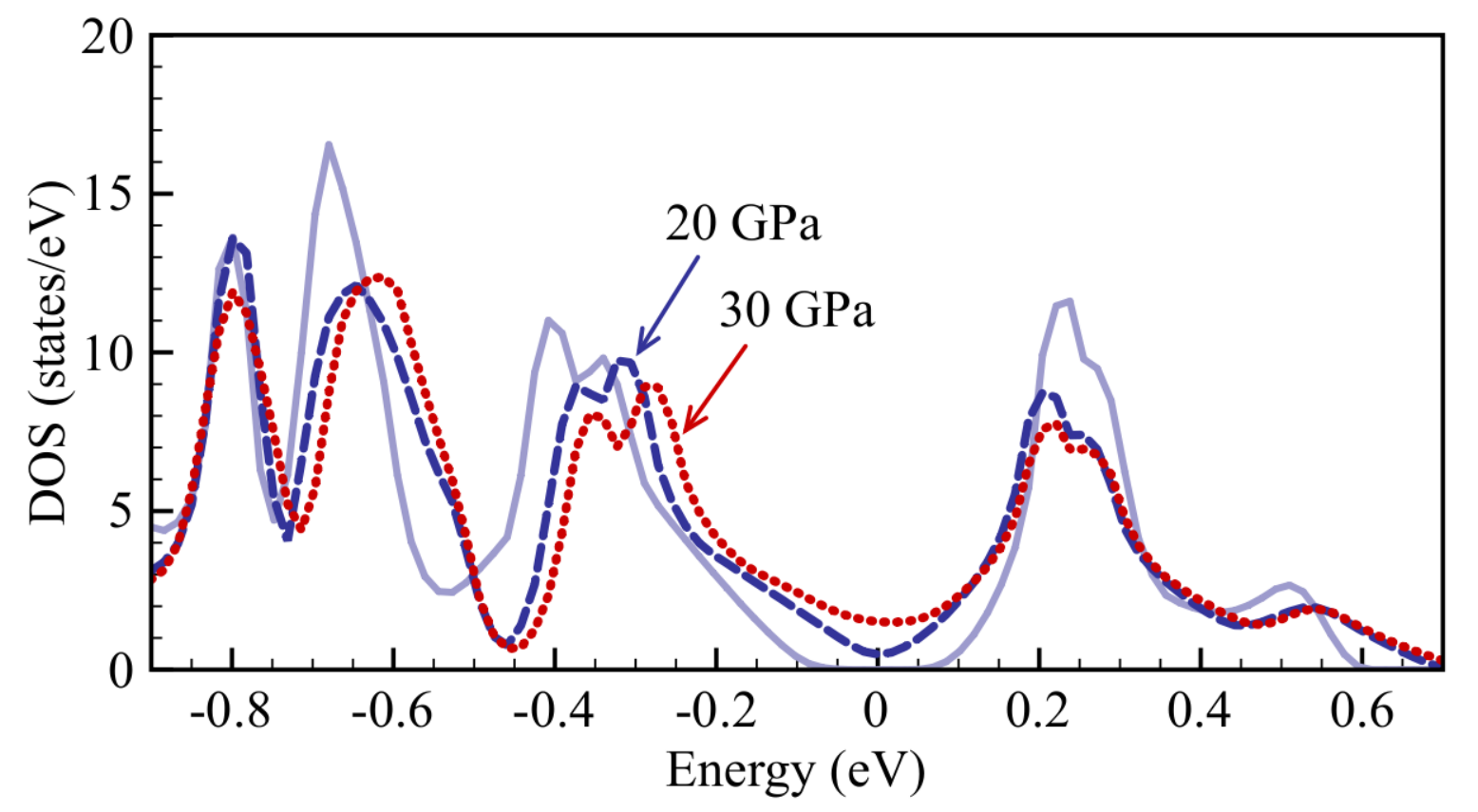}
\caption{(Color online). Total densities of states of Sr$_3$Ir$_2$O$_7$ in the $Bbcb$ structure, calculated at different pressures using the LDA+$U$+SO approach. Solid, dashed and dotted lines correspond to ambient pressure, $P = 20$ GPa and $P = 30$ GPa, respectively.}
\label{dos_comp}
\end{figure}

In our calculations, the insulating state of Sr$_3$Ir$_2$O$_7$ is therefore very sensitive to the value of the pressure. Table \ref{Pressuredata} shows the obtained dependence of the electronic gap on pressure. Above 20 GPa, we find a closure of the electronic gap and a transition to a metallic state. Experimentally, Li \emph{et al.} reported a transition to a nearly metallic state at 13.2 GPa \cite{li2013tuning} with a vanishingly small electronic gap. However, high-pressure transport measurements by Zocco \emph{et al.} \cite{zocco2014persistent} suggest that an electronic gap remains up to at least 30 GPa. The calculated densities of states of Sr$_3$Ir$_{2}$O$_7$ near the Fermi level as a function of pressure are shown in Fig. \ref{dos_comp}. There is a pseudogap state at $P = 20$ GPa; for larger pressures we obtain a metallic solution with non-zero density at the Fermi level.

Another interesting effect evident in our calculations is a suppression of the magnetic moment at high pressure. Analysis of the calculated occupation matrices shows that the strongest contribution to the orbital magnetic moment of Ir comes from the $5d$ states with magnetic quantum number $m_l = \pm$ 1 that correspond to partially-filled, degenerate $xz$ and $yz$ states in the LDA spectrum. This contribution is suppressed under pressure due to strong hybridization with oxygen states. An indication of the hybridization amplification is the LDA spectral function spreading at high pressures (Fig. \ref{dos_amb_high}).

\begin{figure}[!b]
\includegraphics[width=0.9\linewidth]{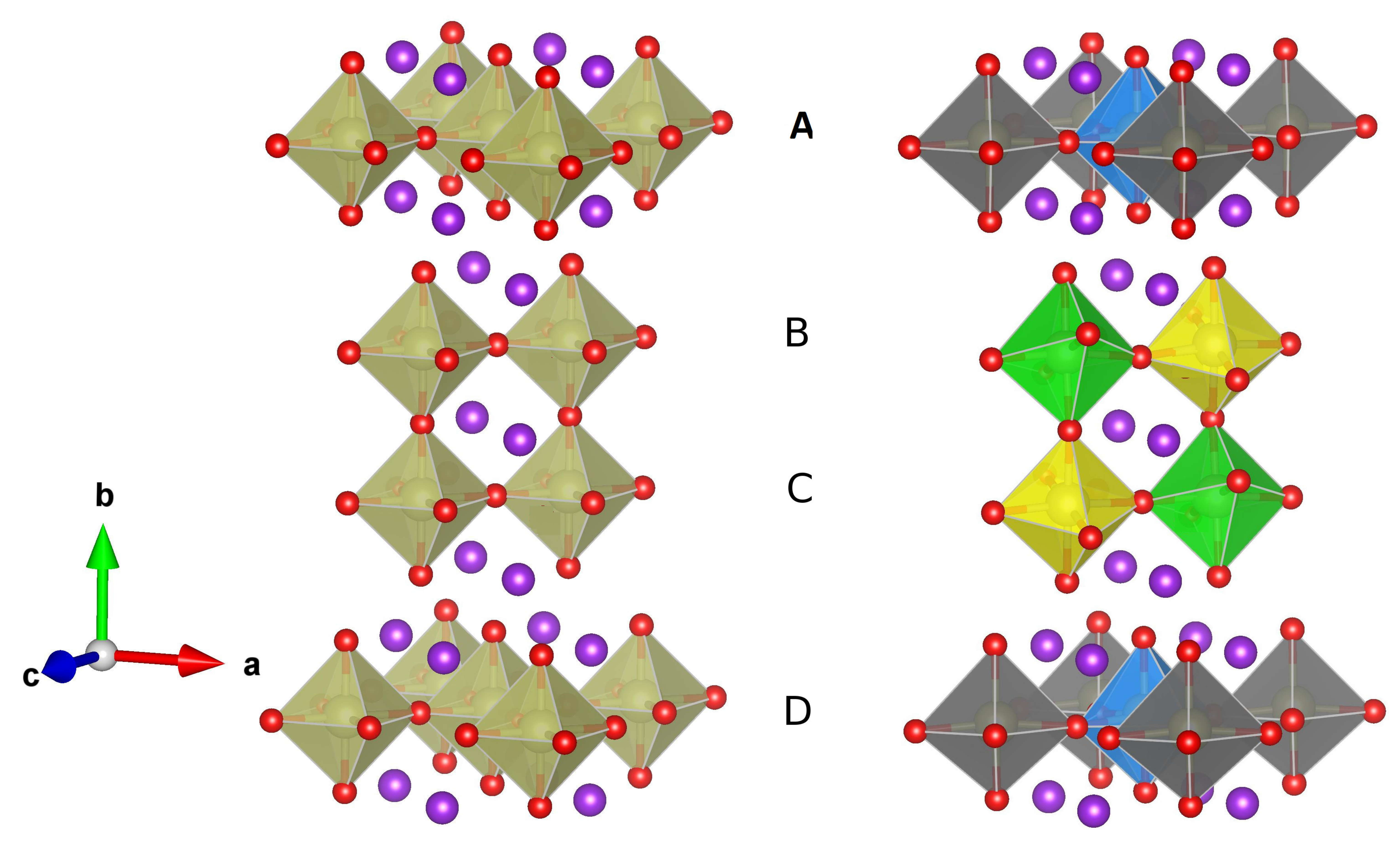}
\caption{(Color online). High-pressure structures of Sr$_{3}$Ir$_2$O$_7$ used in this work. (Left) The structure without rotations of the oxygen octahedra, as determined by experiment. (Right) The optimized structure corresponding to the ground state at high pressure, $P= 61$ GPa. The octahedra belonging to the A (D) and B (C) layers rotate around out-of-plane and in-plane axes, respectively.}
\label{oxygen_octahedra}
\end{figure} 

\subsection{High-pressure phase of Sr$_{3}$Ir$_2$O$_7$}

Analysis of the experimental data established a structural transition to a monoclinic phase at 54 GPa. Since a precise refinement of the oxygen positions from the experimental data was not possible, we explored different arrangements of the oxygen octahedra to optimize the crystal structure of Sr$_{3}$Ir$_2$O$_7$ in the $C2$ phase. We simulated supercells with and without rotations of the oxygen octahedra (Fig. \ref{oxygen_octahedra}). To examine the nature of the structural transition, we also compare our results with a hypothetical $Bbcb$-type structure at 61 GPa, obtained through an extrapolation of the lattice parameters.

\begin{figure}[b]
\includegraphics[width=0.9\linewidth]{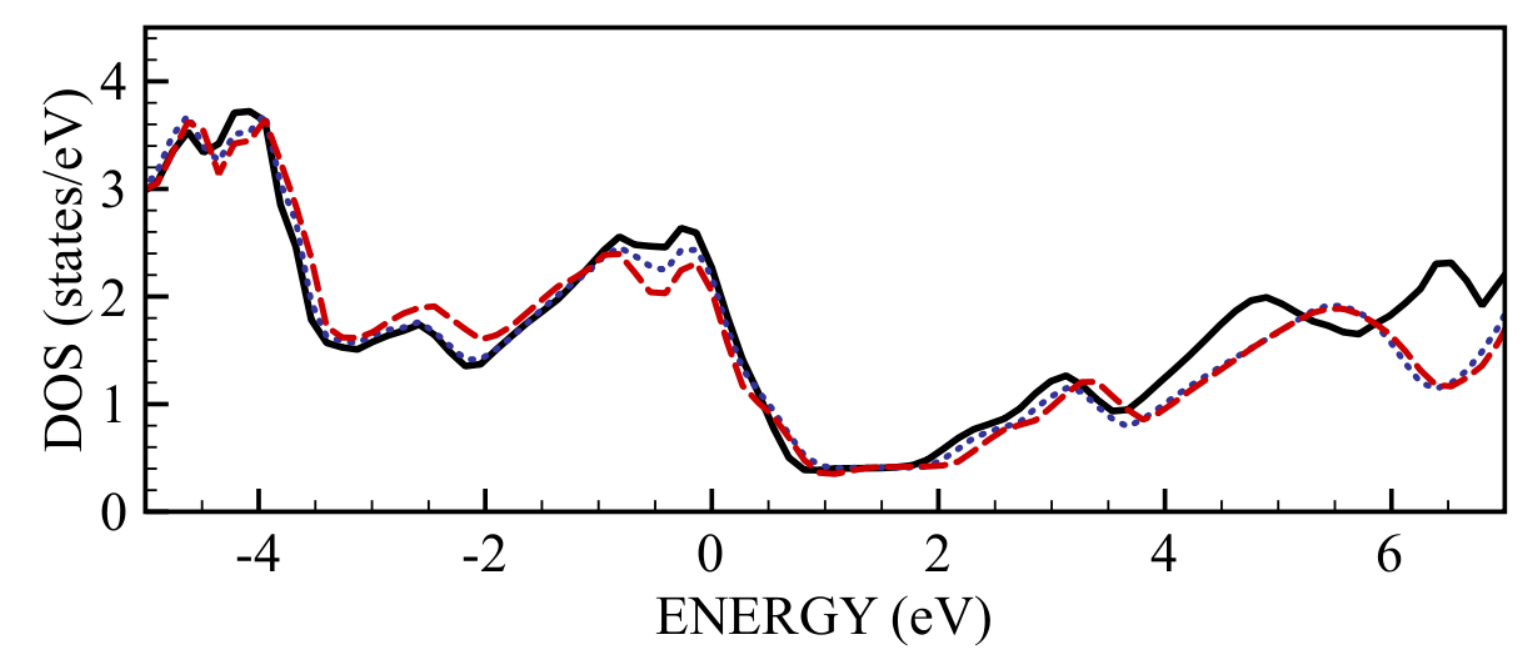}
\caption{(Color online). Comparison of the total densities of states of the refined high-pressure structure of Sr$_3$Ir$_2$O$_7$ without rotation of the oxygen octahedra, obtained from LDA (black solid line), LDA+SO (blue dotted line) and LDA+$U$+SO (red dashed line) calculations. The Fermi level corresponds to zero energy.}
\label{C2LDASOU}
\end{figure}

We first discuss the dependence of the refined structure without oxygen rotations on the choice of the first-principles method. The structures obtained with the LDA, LDA+SO, and LDA+$U$+SO approaches were compared (Table \ref{C2-distance}). In all cases, we used the experimental lattice parameters. A tendency to form nonequivalent in-plane bonds is evident. Importantly, the difference in interatomic distances obtained with LDA, LDA+SO and LDA+U+SO methods is between 10$^{-3}$ \AA \, and 10$^{-2}$ \AA. Such a weak dependence of the optimized structure on the choice of the first-principles method was found with both the full-potential Elk package and the plane-wave pseudopotential approach, realized in the Quantum-Espresso simulation package. The total densities of states shown in Fig. \ref{C2LDASOU} corresponds to a metallic state. 

\begin{table}[t]
\centering
\caption [Bset]{Comparison of Ir-O distances (in \AA) of the experimental and refined high-pressure $C2$ structure of Sr$_{3}$Ir$_2$O$_7$, without rotations of the oxygen octahedra (Fig. \ref{oxygen_octahedra}, left), at 61 GPa. The calculations were performed with the experimental lattice parameters, using different first-principles methods.}
\label {C2-distance}
\begin {tabular}{lcccc}
\hline
  bond  & expt. &  LDA & LDA+SO & LDA+$U$+SO \\
\hline
    Ir-O1 (in-plane) & 1.7545  & 1.7589  & 1.7597  & 1.7562  \\  
    Ir-O2 (in-plane) & 1.7545   & 1.7589   & 1.7597  & 1.7562 \\
    Ir-O3 (in-plane) & 1.8449  &  1.8806   & 1.8811  &  1.8804 \\
    Ir-O4 (in-plane) &  1.8449   & 1.8937   & 1.8974  & 1.8989 \\
    Ir-O5 (out-of-plane) & 1.9175   &  1.9223  & 1.9222  & 1.9189 \\
    Ir-O6 (out-of-plane)  &  1.9175   &  1.9223  & 1.9222  & 1.9189 \\
\hline
\end {tabular}
\end {table}

The LDA solution for the $C2$ structure without rotations of the oxygen octahedra (as determined by experiment; Fig. \ref{oxygen_octahedra}, left) is higher in energy by about 3 eV (per unit cell containing two Ir atoms) than the $Bbcb$ structure at 61 GPa. This largely originates from the more compressed volume of the unit cell of the $C2$ structure (240 \AA$^{3}$), compared to the (extrapolated) volume of the $Bbcb$ structure at 61 GPa (250 \AA$^3$).

\begin{figure}
\includegraphics[width=0.9\linewidth]{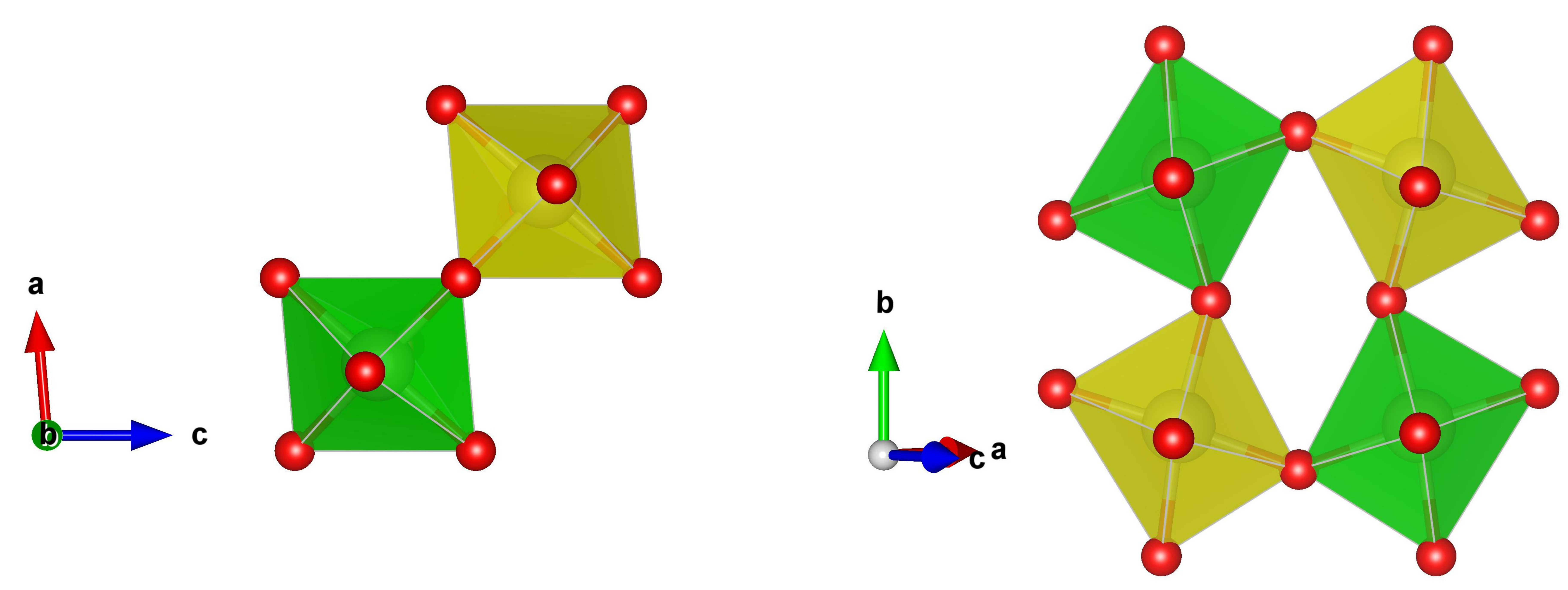}
\caption{(Color online). Top (left panel) and side (right panel) views of the oxygen octahedra belonging to the B(C) plane of Sr$_3$Ir$_2$O$_7$ in the optimized high-pressure structure, as shown in the right panel of Fig. \ref{oxygen_octahedra}.}
\label{oxygen_octahedraBC}
\end{figure}

We now consider the optimized $C2$ structure of Sr$_3$Ir$_2$O$_7$, which adopts complex rotations of the oxygen octahedra (Fig. \ref{oxygen_octahedra}, right). In this case, the LDA calculations revealed that the lowest total energy of the system corresponds to a geometry in which the octahedra from the A and D layers rotate in plane (around the long $b$ axis) by about 9$^{\circ}$. This in-plane rotation is similar to the one observed in the ambient pressure structure. Conversely, the rotation axis for the octahedra from the B and C planes is out of plane (tilting), as shown in Fig. \ref{oxygen_octahedraBC}. We obtained a tilting angle of about 20$^{\circ}$. We also found a tendency for forming short (1.85 \AA) and long (1.94 \AA) bonds within the IrO$_6$ octahedra (Table \ref{hp_distance}).

This optimized high-pressure structure of Sr$_3$Ir$_2$O$_7$ (Fig. \ref{oxygen_octahedra}, right) is 1.4 eV lower in energy (per unit cell containing two Ir atoms) than the experimentally determined high-pressure structure without rotations. We believe that further lowering of the total energy is possible by rotating the IrO$_6$ octahedra in the A and D planes around an in-plane axis. However, this requires an increase in the size of the supercell and optimization of the geometry.

\begin{table}[b]
\centering
\caption [Bset]{Ir-O distances of Sr$_3$Ir$_2$O$_7$ in the optimized high-pressure structure with rotations of the oxygen octahedra, as shown in Fig. \ref{oxygen_octahedraBC}.}
\label {hp_distance}
\begin {tabular}{lcc}
\hline
  plane  & d$_{Ir-O}$ (in-plane) &  d$_{Ir-O}$ (apical) \\
 \hline
    A & 1.85 - 1.94 \AA  & 1.89 - 1.91 \AA  \\  
    B  & 1.87 - 1.92 \AA  & 1.91 - 1.91 \AA  \\
       \hline
   \end {tabular}
\end {table}

\begin{figure}
\includegraphics[width=0.9\linewidth]{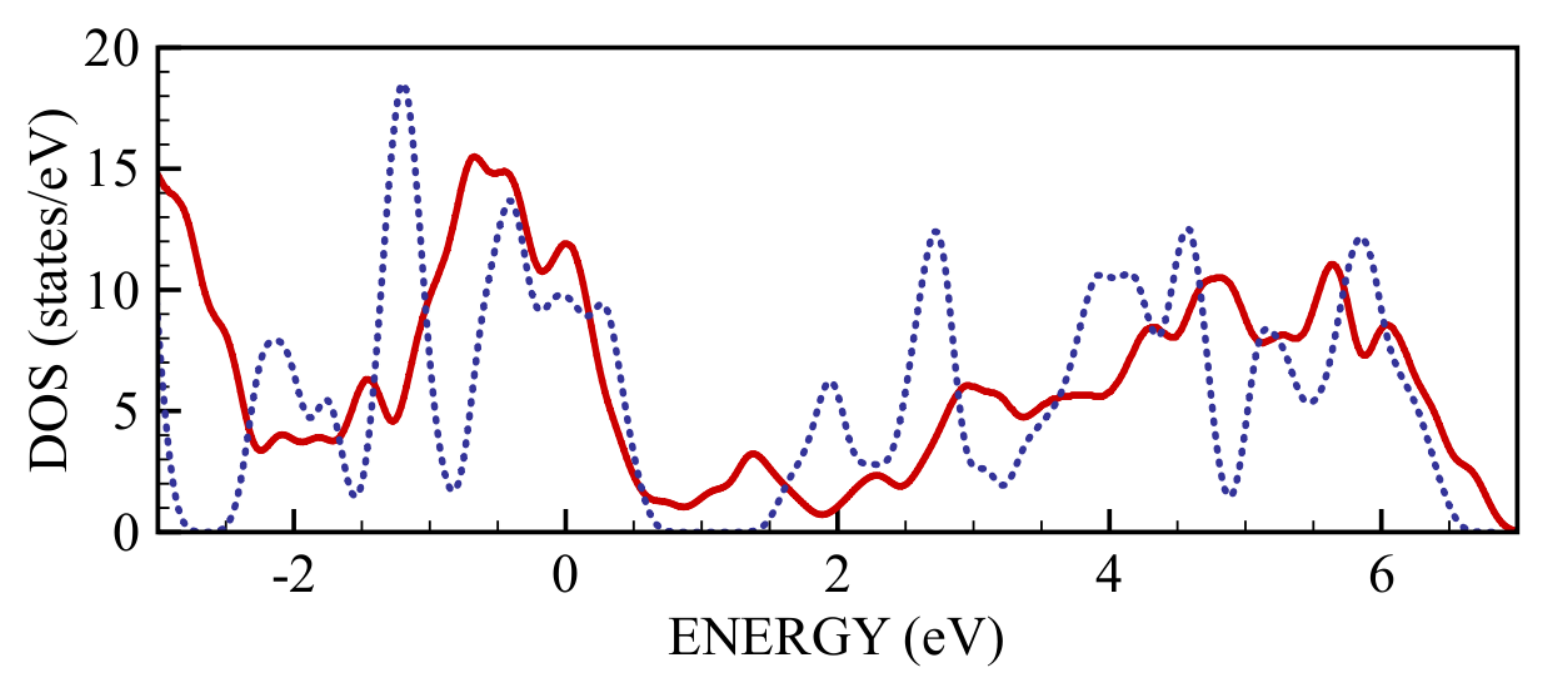}
\caption{(Color online). LDA total densities of states of Sr$_3$Ir$_2$O$_7$ in the compressed $Bbcb$ structure (blue dotted line) and optimized $C2$ structure (red solid line) at $P = 61$ GPa. The Fermi level corresponds to zero energy.}
\label{dos_60Gpa_comp}
\end{figure}

It is important to analyze the difference in the electronic structure of the high-pressure structure of Sr$_3$Ir$_2$O$_7$. Figure \ref{dos_60Gpa_comp} compares the total densities of states of the compressed $Bbcb$ and optimized $C2$ structures, obtained with LDA. The plane-dependent rotation of the oxygen octahedra and compression of the unit cell in the $C2$ structure lead to a complete delocalization of the $5d$ Ir states in Sr$_3$Ir$_2$O$_7$. No electronic gap between the $t_{2g}$ and $e_{g}$ bands remains. 

\begin{figure}
\includegraphics[width=0.9\linewidth]{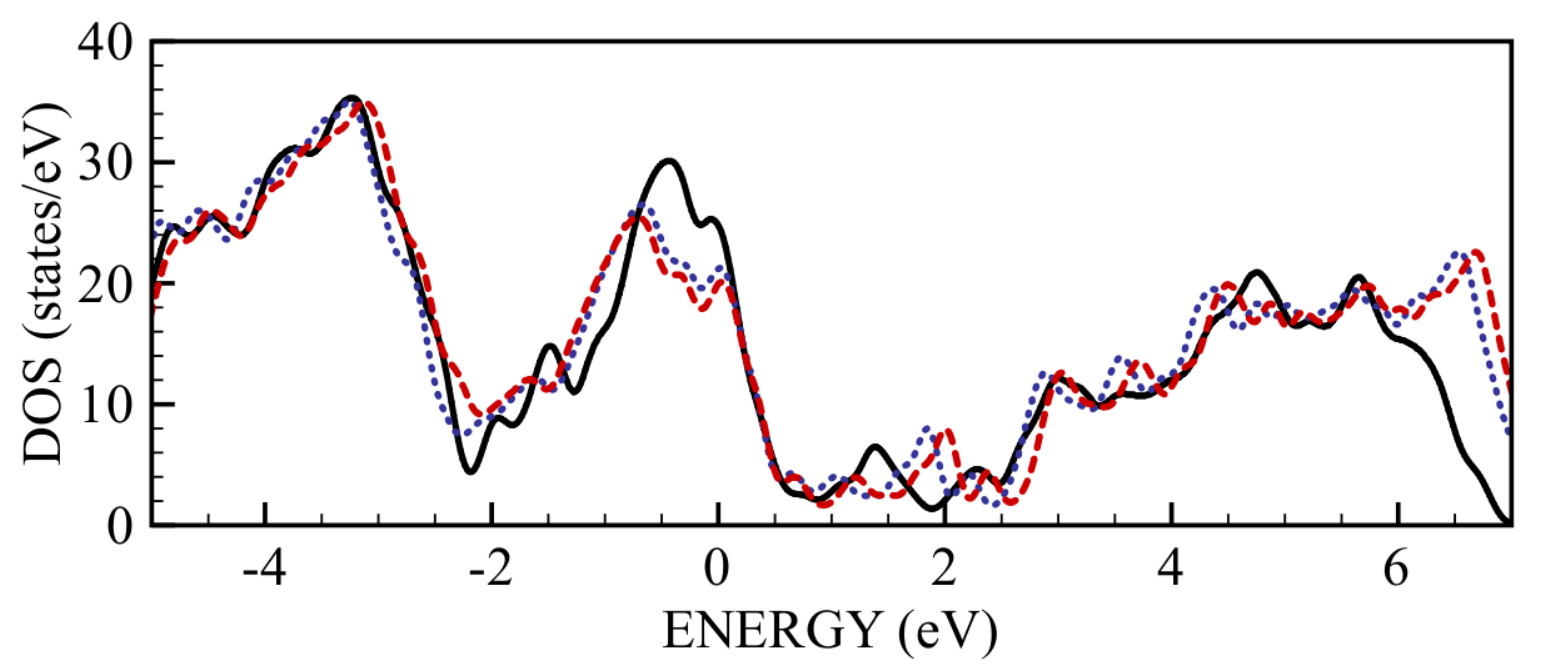}
\caption{(Color online). Comparison of the total densities of states of the optimized $C2$ structure of Sr$_3$Ir$_2$O$_7$ obtained from LDA (black solid line), LDA+SO (blue dotted line) and LDA+$U$+SO (red dashed line) calculations . The Fermi level corresponds to zero energy.}
\label{comp_C2dist}
\end{figure}

Our LDA+SO and LDA+U+SO calculations for an optimized $C2$ structure with a complex rotation of the oxygen octahedra have demonstrated that neither spin-orbit coupling nor on-site Coulomb interaction can significantly affect the electronic structure. The corresponding densities of states are presented in Fig. \ref{comp_C2dist}. The system remains metallic, which still follows from the increased $t_{2g}$ bandwidth. 

\section{Conclusion}

We have studied the structural evolution of Sr$_3$Ir$_2$O$_7$ with powder x-ray diffraction to unprecedented pressures. We observe a novel first-order structural transition at 54 GPa, preceded by a strongly anisotropic compression, characterized by an increasing $c/a$ ratio with pressure. The high-pressure phase could be indexed in a monoclinic symmetry (space group $C2/m$, $C2$, or $Cm$). Rietveld refinement of the diffraction data suggests that the stacking sequence of the perovskite bilayers is altered in the high-pressure structure. A precise refinement of the oxygen positions was, however, not possible due to the relatively weak x-ray scattering signal compared to Ir and Sr atoms.

We used first-principles LDA+$U$+SO calculations to both optimize the structure and examine the electronic properties of Sr$_3$Ir$_2$O$_7$ as a function of pressure. Below the high-pressure structural transition, we find that the anisotropic compression observed by experiment is accompanied by an increase in the in-plane IrO$_6$ octahedral rotation. The simultaneous decrease of the Ir-O bond length however increases the $t_{2g}$ bandwidth and gradually closes the band gap. In our calculations, we find an insulator-metal transition at $\sim$20 GPa. This value is comparable to the transition to a nearly metallic state at 13.2 GPa observed by Li \emph{et al.} \cite{li2013tuning}, but contrasts with measurements by Zocco \emph{et al.} \cite{zocco2014persistent}, which show an electronic gap remaining up to at least 30 GPa.

For the high-pressure phase, structural optimization of the oxygen position resulted in complex in-plane and out-of-plane rotations of the octahedra. This considerably lowered the total energy with respect to the structure obtained from experiment. In the optimized high-pressure structure, the $t_{2g}$ and $e_g$ bands strongly overlap. The shortened Ir-O bond length does not allow an electronic band gap to open, even in the presence of strong spin-orbit coupling and on-site Coulomb interactions. This suggests a robust metallic state for the high-pressure monoclinic phase of Sr$_3$Ir$_2$O$_7$.\\

\begin{acknowledgments}
This work is supported by the UK Engineering and Physical Sciences Research Council, Grant No. EP/J016713/1. H.M.R. acknowledges support from the Swiss National Science Foundation and its Sinergia network Mott Physics Beyond the Heisenberg Model. The work of I.V.S. and V.V.M. is supported by the grant program of the Russian Science Foundation, 14-12-00306. We would like to thank D. Daisenberger, A. Kleppe and H. Wilhelm from beamline I15, Diamond Light Source for technical support.
\end{acknowledgments}


\begin{thebibliography}{33}%
\makeatletter
\providecommand \@ifxundefined [1]{%
 \@ifx{#1\undefined}
}%
\providecommand \@ifnum [1]{%
 \ifnum #1\expandafter \@firstoftwo
 \else \expandafter \@secondoftwo
 \fi
}%
\providecommand \@ifx [1]{%
 \ifx #1\expandafter \@firstoftwo
 \else \expandafter \@secondoftwo
 \fi
}%
\providecommand \natexlab [1]{#1}%
\providecommand \enquote  [1]{``#1''}%
\providecommand \bibnamefont  [1]{#1}%
\providecommand \bibfnamefont [1]{#1}%
\providecommand \citenamefont [1]{#1}%
\providecommand \href@noop [0]{\@secondoftwo}%
\providecommand \href [0]{\begingroup \@sanitize@url \@href}%
\providecommand \@href[1]{\@@startlink{#1}\@@href}%
\providecommand \@@href[1]{\endgroup#1\@@endlink}%
\providecommand \@sanitize@url [0]{\catcode `\\12\catcode `\$12\catcode
  `\&12\catcode `\#12\catcode `\^12\catcode `\_12\catcode `\%12\relax}%
\providecommand \@@startlink[1]{}%
\providecommand \@@endlink[0]{}%
\providecommand \url  [0]{\begingroup\@sanitize@url \@url }%
\providecommand \@url [1]{\endgroup\@href {#1}{\urlprefix }}%
\providecommand \urlprefix  [0]{URL }%
\providecommand \Eprint [0]{\href }%
\providecommand \doibase [0]{http://dx.doi.org/}%
\providecommand \selectlanguage [0]{\@gobble}%
\providecommand \bibinfo  [0]{\@secondoftwo}%
\providecommand \bibfield  [0]{\@secondoftwo}%
\providecommand \translation [1]{[#1]}%
\providecommand \BibitemOpen [0]{}%
\providecommand \bibitemStop [0]{}%
\providecommand \bibitemNoStop [0]{.\EOS\space}%
\providecommand \EOS [0]{\spacefactor3000\relax}%
\providecommand \BibitemShut  [1]{\csname bibitem#1\endcsname}%
\let\auto@bib@innerbib\@empty
%</preamble>
\bibitem [{\citenamefont {Imada}\ \emph {et~al.}(1998)\citenamefont {Imada},
  \citenamefont {Fujimori},\ and\ \citenamefont {Tokura}}]{imada1998metal}%
  \BibitemOpen
  \bibfield  {author} {\bibinfo {author} {\bibfnamefont {M.}~\bibnamefont
  {Imada}}, \bibinfo {author} {\bibfnamefont {A.}~\bibnamefont {Fujimori}}, \
  and\ \bibinfo {author} {\bibfnamefont {Y.}~\bibnamefont {Tokura}},\ }\href
  {\doibase 10.1103/RevModPhys.70.1039} {\bibfield  {journal} {\bibinfo
  {journal} {Rev. Mod. Phys.}\ }\textbf {\bibinfo {volume} {70}},\ \bibinfo
  {pages} {1039} (\bibinfo {year} {1998})}\BibitemShut {NoStop}%
\bibitem [{\citenamefont {Witczak-Krempa}\ \emph {et~al.}(2014)\citenamefont
  {Witczak-Krempa}, \citenamefont {Chen}, \citenamefont {Kim},\ and\
  \citenamefont {Balents}}]{krempa2014correlated}%
  \BibitemOpen
  \bibfield  {author} {\bibinfo {author} {\bibfnamefont {W.}~\bibnamefont
  {Witczak-Krempa}}, \bibinfo {author} {\bibfnamefont {G.}~\bibnamefont
  {Chen}}, \bibinfo {author} {\bibfnamefont {Y.~B.}\ \bibnamefont {Kim}}, \
  and\ \bibinfo {author} {\bibfnamefont {L.}~\bibnamefont {Balents}},\ }\href
  {\doibase 10.1146/annurev-conmatphys-020911-125138} {\bibfield  {journal}
  {\bibinfo  {journal} {Annu. Rev. of Condens. Matter Phys.}\ }\textbf
  {\bibinfo {volume} {5}},\ \bibinfo {pages} {57} (\bibinfo {year}
  {2014})}\BibitemShut {NoStop}%
\bibitem [{\citenamefont {Kim}\ \emph {et~al.}(2008)\citenamefont {Kim},
  \citenamefont {Jin}, \citenamefont {Moon}, \citenamefont {Kim}, \citenamefont
  {Park}, \citenamefont {Leem}, \citenamefont {Yu}, \citenamefont {Noh},
  \citenamefont {Kim}, \citenamefont {Oh}, \citenamefont {Park}, \citenamefont
  {Durairaj}, \citenamefont {Cao},\ and\ \citenamefont
  {Rotenberg}}]{kim2008novel}%
  \BibitemOpen
  \bibfield  {author} {\bibinfo {author} {\bibfnamefont {B.~J.}\ \bibnamefont
  {Kim}}, \bibinfo {author} {\bibfnamefont {H.}~\bibnamefont {Jin}}, \bibinfo
  {author} {\bibfnamefont {S.~J.}\ \bibnamefont {Moon}}, \bibinfo {author}
  {\bibfnamefont {J.-Y.}\ \bibnamefont {Kim}}, \bibinfo {author} {\bibfnamefont
  {B.-G.}\ \bibnamefont {Park}}, \bibinfo {author} {\bibfnamefont {C.~S.}\
  \bibnamefont {Leem}}, \bibinfo {author} {\bibfnamefont {J.}~\bibnamefont
  {Yu}}, \bibinfo {author} {\bibfnamefont {T.~W.}\ \bibnamefont {Noh}},
  \bibinfo {author} {\bibfnamefont {C.}~\bibnamefont {Kim}}, \bibinfo {author}
  {\bibfnamefont {S.-J.}\ \bibnamefont {Oh}}, \bibinfo {author} {\bibfnamefont
  {J.-H.}\ \bibnamefont {Park}}, \bibinfo {author} {\bibfnamefont
  {V.}~\bibnamefont {Durairaj}}, \bibinfo {author} {\bibfnamefont
  {G.}~\bibnamefont {Cao}}, \ and\ \bibinfo {author} {\bibfnamefont
  {E.}~\bibnamefont {Rotenberg}},\ }\href {\doibase
  10.1103/PhysRevLett.101.076402} {\bibfield  {journal} {\bibinfo  {journal}
  {Phys. Rev. Lett.}\ }\textbf {\bibinfo {volume} {101}},\ \bibinfo {pages}
  {076402} (\bibinfo {year} {2008})}\BibitemShut {NoStop}%
\bibitem [{\citenamefont {Moon}\ \emph {et~al.}(2008)\citenamefont {Moon},
  \citenamefont {Jin}, \citenamefont {Kim}, \citenamefont {Choi}, \citenamefont
  {Lee}, \citenamefont {Yu}, \citenamefont {Cao}, \citenamefont {Sumi},
  \citenamefont {Funakubo}, \citenamefont {Bernhard},\ and\ \citenamefont
  {Noh}}]{moon2008dimensionality}%
  \BibitemOpen
  \bibfield  {author} {\bibinfo {author} {\bibfnamefont {S.~J.}\ \bibnamefont
  {Moon}}, \bibinfo {author} {\bibfnamefont {H.}~\bibnamefont {Jin}}, \bibinfo
  {author} {\bibfnamefont {K.~W.}\ \bibnamefont {Kim}}, \bibinfo {author}
  {\bibfnamefont {W.~S.}\ \bibnamefont {Choi}}, \bibinfo {author}
  {\bibfnamefont {Y.~S.}\ \bibnamefont {Lee}}, \bibinfo {author} {\bibfnamefont
  {J.}~\bibnamefont {Yu}}, \bibinfo {author} {\bibfnamefont {G.}~\bibnamefont
  {Cao}}, \bibinfo {author} {\bibfnamefont {A.}~\bibnamefont {Sumi}}, \bibinfo
  {author} {\bibfnamefont {H.}~\bibnamefont {Funakubo}}, \bibinfo {author}
  {\bibfnamefont {C.}~\bibnamefont {Bernhard}}, \ and\ \bibinfo {author}
  {\bibfnamefont {T.~W.}\ \bibnamefont {Noh}},\ }\href {\doibase
  10.1103/PhysRevLett.101.226402} {\bibfield  {journal} {\bibinfo  {journal}
  {Phys. Rev. Lett.}\ }\textbf {\bibinfo {volume} {101}},\ \bibinfo {pages}
  {226402} (\bibinfo {year} {2008})}\BibitemShut {NoStop}%
\bibitem [{\citenamefont {Arita}\ \emph {et~al.}(2012)\citenamefont {Arita},
  \citenamefont {Kune\ifmmode~\check{s}\else \v{s}\fi{}}, \citenamefont
  {Kozhevnikov}, \citenamefont {Eguiluz},\ and\ \citenamefont
  {Imada}}]{arita2012ab}%
  \BibitemOpen
  \bibfield  {author} {\bibinfo {author} {\bibfnamefont {R.}~\bibnamefont
  {Arita}}, \bibinfo {author} {\bibfnamefont {J.}~\bibnamefont
  {Kune\ifmmode~\check{s}\else \v{s}\fi{}}}, \bibinfo {author} {\bibfnamefont
  {A.~V.}\ \bibnamefont {Kozhevnikov}}, \bibinfo {author} {\bibfnamefont
  {A.~G.}\ \bibnamefont {Eguiluz}}, \ and\ \bibinfo {author} {\bibfnamefont
  {M.}~\bibnamefont {Imada}},\ }\href {\doibase 10.1103/PhysRevLett.108.086403}
  {\bibfield  {journal} {\bibinfo  {journal} {Phys. Rev. Lett.}\ }\textbf
  {\bibinfo {volume} {108}},\ \bibinfo {pages} {086403} (\bibinfo {year}
  {2012})}\BibitemShut {NoStop}%
\bibitem [{\citenamefont {Kim}\ \emph {et~al.}(2009)\citenamefont {Kim},
  \citenamefont {Ohsumi}, \citenamefont {Komesu}, \citenamefont {Sakai},
  \citenamefont {Morita}, \citenamefont {Takagi},\ and\ \citenamefont
  {Arima}}]{kim2009phase}%
  \BibitemOpen
  \bibfield  {author} {\bibinfo {author} {\bibfnamefont {B.~J.}\ \bibnamefont
  {Kim}}, \bibinfo {author} {\bibfnamefont {H.}~\bibnamefont {Ohsumi}},
  \bibinfo {author} {\bibfnamefont {T.}~\bibnamefont {Komesu}}, \bibinfo
  {author} {\bibfnamefont {S.}~\bibnamefont {Sakai}}, \bibinfo {author}
  {\bibfnamefont {T.}~\bibnamefont {Morita}}, \bibinfo {author} {\bibfnamefont
  {H.}~\bibnamefont {Takagi}}, \ and\ \bibinfo {author} {\bibfnamefont
  {T.}~\bibnamefont {Arima}},\ }\href {\doibase 10.1126/science.1167106}
  {\bibfield  {journal} {\bibinfo  {journal} {Science}\ }\textbf {\bibinfo
  {volume} {323}},\ \bibinfo {pages} {1329} (\bibinfo {year}
  {2009})}\BibitemShut {NoStop}%
\bibitem [{\citenamefont {Jackeli}\ and\ \citenamefont
  {Khaliullin}(2009)}]{jackeli2009mott}%
  \BibitemOpen
  \bibfield  {author} {\bibinfo {author} {\bibfnamefont {G.}~\bibnamefont
  {Jackeli}}\ and\ \bibinfo {author} {\bibfnamefont {G.}~\bibnamefont
  {Khaliullin}},\ }\href {\doibase 10.1103/PhysRevLett.102.017205} {\bibfield
  {journal} {\bibinfo  {journal} {Phys. Rev. Lett.}\ }\textbf {\bibinfo
  {volume} {102}},\ \bibinfo {pages} {017205} (\bibinfo {year}
  {2009})}\BibitemShut {NoStop}%
\bibitem [{\citenamefont {Cao}\ \emph {et~al.}(2002)\citenamefont {Cao},
  \citenamefont {Xin}, \citenamefont {Alexander}, \citenamefont {Crow},
  \citenamefont {Schlottmann}, \citenamefont {Crawford}, \citenamefont
  {Harlow},\ and\ \citenamefont {Marshall}}]{cao2002anomalous}%
  \BibitemOpen
  \bibfield  {author} {\bibinfo {author} {\bibfnamefont {G.}~\bibnamefont
  {Cao}}, \bibinfo {author} {\bibfnamefont {Y.}~\bibnamefont {Xin}}, \bibinfo
  {author} {\bibfnamefont {C.~S.}\ \bibnamefont {Alexander}}, \bibinfo {author}
  {\bibfnamefont {J.~E.}\ \bibnamefont {Crow}}, \bibinfo {author}
  {\bibfnamefont {P.}~\bibnamefont {Schlottmann}}, \bibinfo {author}
  {\bibfnamefont {M.~K.}\ \bibnamefont {Crawford}}, \bibinfo {author}
  {\bibfnamefont {R.~L.}\ \bibnamefont {Harlow}}, \ and\ \bibinfo {author}
  {\bibfnamefont {W.}~\bibnamefont {Marshall}},\ }\href {\doibase
  10.1103/PhysRevB.66.214412} {\bibfield  {journal} {\bibinfo  {journal} {Phys.
  Rev. B}\ }\textbf {\bibinfo {volume} {66}},\ \bibinfo {pages} {214412}
  (\bibinfo {year} {2002})}\BibitemShut {NoStop}%
\bibitem [{\citenamefont {Dhital}\ \emph {et~al.}(2012)\citenamefont {Dhital},
  \citenamefont {Khadka}, \citenamefont {Yamani}, \citenamefont {de~la Cruz},
  \citenamefont {Hogan}, \citenamefont {Disseler}, \citenamefont {Pokharel},
  \citenamefont {Lukas}, \citenamefont {Tian}, \citenamefont {Opeil},
  \citenamefont {Wang},\ and\ \citenamefont {Wilson}}]{dhital2012spin}%
  \BibitemOpen
  \bibfield  {author} {\bibinfo {author} {\bibfnamefont {C.}~\bibnamefont
  {Dhital}}, \bibinfo {author} {\bibfnamefont {S.}~\bibnamefont {Khadka}},
  \bibinfo {author} {\bibfnamefont {Z.}~\bibnamefont {Yamani}}, \bibinfo
  {author} {\bibfnamefont {C.}~\bibnamefont {de~la Cruz}}, \bibinfo {author}
  {\bibfnamefont {T.~C.}\ \bibnamefont {Hogan}}, \bibinfo {author}
  {\bibfnamefont {S.~M.}\ \bibnamefont {Disseler}}, \bibinfo {author}
  {\bibfnamefont {M.}~\bibnamefont {Pokharel}}, \bibinfo {author}
  {\bibfnamefont {K.~C.}\ \bibnamefont {Lukas}}, \bibinfo {author}
  {\bibfnamefont {W.}~\bibnamefont {Tian}}, \bibinfo {author} {\bibfnamefont
  {C.~P.}\ \bibnamefont {Opeil}}, \bibinfo {author} {\bibfnamefont
  {Z.}~\bibnamefont {Wang}}, \ and\ \bibinfo {author} {\bibfnamefont {S.~D.}\
  \bibnamefont {Wilson}},\ }\href {\doibase 10.1103/PhysRevB.86.100401}
  {\bibfield  {journal} {\bibinfo  {journal} {Phys. Rev. B}\ }\textbf {\bibinfo
  {volume} {86}},\ \bibinfo {pages} {100401} (\bibinfo {year}
  {2012})}\BibitemShut {NoStop}%
\bibitem [{\citenamefont {Hsieh}\ \emph {et~al.}(2012)\citenamefont {Hsieh},
  \citenamefont {Mahmood}, \citenamefont {Torchinsky}, \citenamefont {Cao},\
  and\ \citenamefont {Gedik}}]{hsieh2012observation}%
  \BibitemOpen
  \bibfield  {author} {\bibinfo {author} {\bibfnamefont {D.}~\bibnamefont
  {Hsieh}}, \bibinfo {author} {\bibfnamefont {F.}~\bibnamefont {Mahmood}},
  \bibinfo {author} {\bibfnamefont {D.~H.}\ \bibnamefont {Torchinsky}},
  \bibinfo {author} {\bibfnamefont {G.}~\bibnamefont {Cao}}, \ and\ \bibinfo
  {author} {\bibfnamefont {N.}~\bibnamefont {Gedik}},\ }\href {\doibase
  10.1103/PhysRevB.86.035128} {\bibfield  {journal} {\bibinfo  {journal} {Phys.
  Rev. B}\ }\textbf {\bibinfo {volume} {86}},\ \bibinfo {pages} {035128}
  (\bibinfo {year} {2012})}\BibitemShut {NoStop}%
\bibitem [{\citenamefont {Watanabe}\ \emph {et~al.}(2014)\citenamefont
  {Watanabe}, \citenamefont {Shirakawa},\ and\ \citenamefont
  {Yunoki}}]{watanabe2014theoretical}%
  \BibitemOpen
  \bibfield  {author} {\bibinfo {author} {\bibfnamefont {H.}~\bibnamefont
  {Watanabe}}, \bibinfo {author} {\bibfnamefont {T.}~\bibnamefont {Shirakawa}},
  \ and\ \bibinfo {author} {\bibfnamefont {S.}~\bibnamefont {Yunoki}},\ }\href
  {\doibase 10.1103/PhysRevB.89.165115} {\bibfield  {journal} {\bibinfo
  {journal} {Phys. Rev. B}\ }\textbf {\bibinfo {volume} {89}},\ \bibinfo
  {pages} {165115} (\bibinfo {year} {2014})}\BibitemShut {NoStop}%
\bibitem [{\citenamefont {Kim}\ \emph {et~al.}(2012)\citenamefont {Kim},
  \citenamefont {Choi}, \citenamefont {Kim}, \citenamefont {Mitchell},
  \citenamefont {Jackeli}, \citenamefont {Daghofer}, \citenamefont {van~den
  Brink}, \citenamefont {Khaliullin},\ and\ \citenamefont
  {Kim}}]{kim2012dimensionality}%
  \BibitemOpen
  \bibfield  {author} {\bibinfo {author} {\bibfnamefont {J.~W.}\ \bibnamefont
  {Kim}}, \bibinfo {author} {\bibfnamefont {Y.}~\bibnamefont {Choi}}, \bibinfo
  {author} {\bibfnamefont {J.}~\bibnamefont {Kim}}, \bibinfo {author}
  {\bibfnamefont {J.~F.}\ \bibnamefont {Mitchell}}, \bibinfo {author}
  {\bibfnamefont {G.}~\bibnamefont {Jackeli}}, \bibinfo {author} {\bibfnamefont
  {M.}~\bibnamefont {Daghofer}}, \bibinfo {author} {\bibfnamefont
  {J.}~\bibnamefont {van~den Brink}}, \bibinfo {author} {\bibfnamefont
  {G.}~\bibnamefont {Khaliullin}}, \ and\ \bibinfo {author} {\bibfnamefont
  {B.~J.}\ \bibnamefont {Kim}},\ }\href {\doibase
  10.1103/PhysRevLett.109.037204} {\bibfield  {journal} {\bibinfo  {journal}
  {Phys. Rev. Lett.}\ }\textbf {\bibinfo {volume} {109}},\ \bibinfo {pages}
  {037204} (\bibinfo {year} {2012})}\BibitemShut {NoStop}%
\bibitem [{\citenamefont {Boseggia}\ \emph {et~al.}(2012)\citenamefont
  {Boseggia}, \citenamefont {Springell}, \citenamefont {Walker}, \citenamefont
  {Boothroyd}, \citenamefont {Prabhakaran}, \citenamefont {Wermeille},
  \citenamefont {Bouchenoire}, \citenamefont {Collins},\ and\ \citenamefont
  {McMorrow}}]{boseggia2012antiferromagnetic}%
  \BibitemOpen
  \bibfield  {author} {\bibinfo {author} {\bibfnamefont {S.}~\bibnamefont
  {Boseggia}}, \bibinfo {author} {\bibfnamefont {R.}~\bibnamefont {Springell}},
  \bibinfo {author} {\bibfnamefont {H.~C.}\ \bibnamefont {Walker}}, \bibinfo
  {author} {\bibfnamefont {A.~T.}\ \bibnamefont {Boothroyd}}, \bibinfo {author}
  {\bibfnamefont {D.}~\bibnamefont {Prabhakaran}}, \bibinfo {author}
  {\bibfnamefont {D.}~\bibnamefont {Wermeille}}, \bibinfo {author}
  {\bibfnamefont {L.}~\bibnamefont {Bouchenoire}}, \bibinfo {author}
  {\bibfnamefont {S.~P.}\ \bibnamefont {Collins}}, \ and\ \bibinfo {author}
  {\bibfnamefont {D.~F.}\ \bibnamefont {McMorrow}},\ }\href {\doibase
  10.1103/PhysRevB.85.184432} {\bibfield  {journal} {\bibinfo  {journal} {Phys.
  Rev. B}\ }\textbf {\bibinfo {volume} {85}},\ \bibinfo {pages} {184432}
  (\bibinfo {year} {2012})}\BibitemShut {NoStop}%
\bibitem [{\citenamefont {Zhang}\ \emph {et~al.}(2013)\citenamefont {Zhang},
  \citenamefont {Haule},\ and\ \citenamefont
  {Vanderbilt}}]{zhang2013effective}%
  \BibitemOpen
  \bibfield  {author} {\bibinfo {author} {\bibfnamefont {H.}~\bibnamefont
  {Zhang}}, \bibinfo {author} {\bibfnamefont {K.}~\bibnamefont {Haule}}, \ and\
  \bibinfo {author} {\bibfnamefont {D.}~\bibnamefont {Vanderbilt}},\ }\href
  {\doibase 10.1103/PhysRevLett.111.246402} {\bibfield  {journal} {\bibinfo
  {journal} {Phys. Rev. Lett.}\ }\textbf {\bibinfo {volume} {111}},\ \bibinfo
  {pages} {246402} (\bibinfo {year} {2013})}\BibitemShut {NoStop}%
\bibitem [{\citenamefont {Okada}\ \emph {et~al.}(2013)\citenamefont {Okada},
  \citenamefont {Walkup}, \citenamefont {Lin}, \citenamefont {Dhital},
  \citenamefont {Chang}, \citenamefont {Khadka}, \citenamefont {Zhou},
  \citenamefont {Jeng}, \citenamefont {Paranjape}, \citenamefont {Bansil},
  \citenamefont {Wang}, \citenamefont {Wilson},\ and\ \citenamefont
  {Madhavan}}]{okada2013imaging}%
  \BibitemOpen
  \bibfield  {author} {\bibinfo {author} {\bibfnamefont {Y.}~\bibnamefont
  {Okada}}, \bibinfo {author} {\bibfnamefont {D.}~\bibnamefont {Walkup}},
  \bibinfo {author} {\bibfnamefont {H.}~\bibnamefont {Lin}}, \bibinfo {author}
  {\bibfnamefont {C.}~\bibnamefont {Dhital}}, \bibinfo {author} {\bibfnamefont
  {T.-R.}\ \bibnamefont {Chang}}, \bibinfo {author} {\bibfnamefont
  {S.}~\bibnamefont {Khadka}}, \bibinfo {author} {\bibfnamefont
  {W.}~\bibnamefont {Zhou}}, \bibinfo {author} {\bibfnamefont {H.-T.}\
  \bibnamefont {Jeng}}, \bibinfo {author} {\bibfnamefont {M.}~\bibnamefont
  {Paranjape}}, \bibinfo {author} {\bibfnamefont {A.}~\bibnamefont {Bansil}},
  \bibinfo {author} {\bibfnamefont {Z.}~\bibnamefont {Wang}}, \bibinfo {author}
  {\bibfnamefont {S.~D.}\ \bibnamefont {Wilson}}, \ and\ \bibinfo {author}
  {\bibfnamefont {V.}~\bibnamefont {Madhavan}},\ }\href
  {http://dx.doi.org/10.1038/nmat3653} {\bibfield  {journal} {\bibinfo
  {journal} {Nat. Mater.}\ }\textbf {\bibinfo {volume} {12}},\ \bibinfo {pages}
  {707} (\bibinfo {year} {2013})}\BibitemShut {NoStop}%
\bibitem [{\citenamefont {Li}\ \emph {et~al.}(2013)\citenamefont {Li},
  \citenamefont {Kong}, \citenamefont {Qi}, \citenamefont {Jin}, \citenamefont
  {Yuan}, \citenamefont {DeLong}, \citenamefont {Schlottmann},\ and\
  \citenamefont {Cao}}]{li2013tuning}%
  \BibitemOpen
  \bibfield  {author} {\bibinfo {author} {\bibfnamefont {L.}~\bibnamefont
  {Li}}, \bibinfo {author} {\bibfnamefont {P.~P.}\ \bibnamefont {Kong}},
  \bibinfo {author} {\bibfnamefont {T.~F.}\ \bibnamefont {Qi}}, \bibinfo
  {author} {\bibfnamefont {C.~Q.}\ \bibnamefont {Jin}}, \bibinfo {author}
  {\bibfnamefont {S.~J.}\ \bibnamefont {Yuan}}, \bibinfo {author}
  {\bibfnamefont {L.~E.}\ \bibnamefont {DeLong}}, \bibinfo {author}
  {\bibfnamefont {P.}~\bibnamefont {Schlottmann}}, \ and\ \bibinfo {author}
  {\bibfnamefont {G.}~\bibnamefont {Cao}},\ }\href {\doibase
  10.1103/PhysRevB.87.235127} {\bibfield  {journal} {\bibinfo  {journal} {Phys.
  Rev. B}\ }\textbf {\bibinfo {volume} {87}},\ \bibinfo {pages} {235127}
  (\bibinfo {year} {2013})}\BibitemShut {NoStop}%
\bibitem [{\citenamefont {de~la Torre}\ \emph {et~al.}(2014)\citenamefont
  {de~la Torre}, \citenamefont {Hunter}, \citenamefont {Subedi}, \citenamefont
  {McKeown~Walker}, \citenamefont {Tamai}, \citenamefont {Kim}, \citenamefont
  {Hoesch}, \citenamefont {Perry}, \citenamefont {Georges},\ and\ \citenamefont
  {Baumberger}}]{de2014coherent}%
  \BibitemOpen
  \bibfield  {author} {\bibinfo {author} {\bibfnamefont {A.}~\bibnamefont
  {de~la Torre}}, \bibinfo {author} {\bibfnamefont {E.~C.}\ \bibnamefont
  {Hunter}}, \bibinfo {author} {\bibfnamefont {A.}~\bibnamefont {Subedi}},
  \bibinfo {author} {\bibfnamefont {S.}~\bibnamefont {McKeown~Walker}},
  \bibinfo {author} {\bibfnamefont {A.}~\bibnamefont {Tamai}}, \bibinfo
  {author} {\bibfnamefont {T.~K.}\ \bibnamefont {Kim}}, \bibinfo {author}
  {\bibfnamefont {M.}~\bibnamefont {Hoesch}}, \bibinfo {author} {\bibfnamefont
  {R.~S.}\ \bibnamefont {Perry}}, \bibinfo {author} {\bibfnamefont
  {A.}~\bibnamefont {Georges}}, \ and\ \bibinfo {author} {\bibfnamefont
  {F.}~\bibnamefont {Baumberger}},\ }\href {\doibase
  10.1103/PhysRevLett.113.256402} {\bibfield  {journal} {\bibinfo  {journal}
  {Phys. Rev. Lett.}\ }\textbf {\bibinfo {volume} {113}},\ \bibinfo {pages}
  {256402} (\bibinfo {year} {2014})}\BibitemShut {NoStop}%
\bibitem [{\citenamefont {He}\ \emph {et~al.}(2015)\citenamefont {He},
  \citenamefont {Hogan}, \citenamefont {Mion}, \citenamefont {Hafiz},
  \citenamefont {He}, \citenamefont {Denlinger}, \citenamefont {Mo},
  \citenamefont {Dhital}, \citenamefont {Chen}, \citenamefont {Lin},
  \citenamefont {Zhang}, \citenamefont {Hashimoto}, \citenamefont {Pan},
  \citenamefont {Lu}, \citenamefont {Arita}, \citenamefont {Shimada},
  \citenamefont {Markiewicz}, \citenamefont {Wang}, \citenamefont {Kempa},
  \citenamefont {Naughton}, \citenamefont {Bansil}, \citenamefont {Wilson},\
  and\ \citenamefont {He}}]{he2015spectroscopic}%
  \BibitemOpen
  \bibfield  {author} {\bibinfo {author} {\bibfnamefont {J.}~\bibnamefont
  {He}}, \bibinfo {author} {\bibfnamefont {T.}~\bibnamefont {Hogan}}, \bibinfo
  {author} {\bibfnamefont {T.~R.}\ \bibnamefont {Mion}}, \bibinfo {author}
  {\bibfnamefont {H.}~\bibnamefont {Hafiz}}, \bibinfo {author} {\bibfnamefont
  {Y.}~\bibnamefont {He}}, \bibinfo {author} {\bibfnamefont {J.~D.}\
  \bibnamefont {Denlinger}}, \bibinfo {author} {\bibfnamefont {S.-K.}\
  \bibnamefont {Mo}}, \bibinfo {author} {\bibfnamefont {C.}~\bibnamefont
  {Dhital}}, \bibinfo {author} {\bibfnamefont {X.}~\bibnamefont {Chen}},
  \bibinfo {author} {\bibfnamefont {Q.}~\bibnamefont {Lin}}, \bibinfo {author}
  {\bibfnamefont {Y.}~\bibnamefont {Zhang}}, \bibinfo {author} {\bibfnamefont
  {M.}~\bibnamefont {Hashimoto}}, \bibinfo {author} {\bibfnamefont
  {H.}~\bibnamefont {Pan}}, \bibinfo {author} {\bibfnamefont {D.~H.}\
  \bibnamefont {Lu}}, \bibinfo {author} {\bibfnamefont {M.}~\bibnamefont
  {Arita}}, \bibinfo {author} {\bibfnamefont {K.}~\bibnamefont {Shimada}},
  \bibinfo {author} {\bibfnamefont {R.~S.}\ \bibnamefont {Markiewicz}},
  \bibinfo {author} {\bibfnamefont {Z.}~\bibnamefont {Wang}}, \bibinfo {author}
  {\bibfnamefont {K.}~\bibnamefont {Kempa}}, \bibinfo {author} {\bibfnamefont
  {M.~J.}\ \bibnamefont {Naughton}}, \bibinfo {author} {\bibfnamefont
  {A.}~\bibnamefont {Bansil}}, \bibinfo {author} {\bibfnamefont {S.~D.}\
  \bibnamefont {Wilson}}, \ and\ \bibinfo {author} {\bibfnamefont {R.-H.}\
  \bibnamefont {He}},\ }\href {http://dx.doi.org/10.1038/nmat4273} {\bibfield
  {journal} {\bibinfo  {journal} {Nat. Mater.}\ }\textbf {\bibinfo {volume}
  {14}},\ \bibinfo {pages} {577} (\bibinfo {year} {2015})}\BibitemShut
  {NoStop}%
\bibitem [{\citenamefont {Zhao}\ \emph {et~al.}(2014)\citenamefont {Zhao},
  \citenamefont {Wang}, \citenamefont {Qi}, \citenamefont {Zeng}, \citenamefont
  {Hirai}, \citenamefont {Kong}, \citenamefont {Li}, \citenamefont {Park},
  \citenamefont {Yuan}, \citenamefont {Jin}, \citenamefont {Cao},\ and\
  \citenamefont {Mao}}]{zhao2014pressure}%
  \BibitemOpen
  \bibfield  {author} {\bibinfo {author} {\bibfnamefont {Z.}~\bibnamefont
  {Zhao}}, \bibinfo {author} {\bibfnamefont {S.}~\bibnamefont {Wang}}, \bibinfo
  {author} {\bibfnamefont {T.~F.}\ \bibnamefont {Qi}}, \bibinfo {author}
  {\bibfnamefont {Q.}~\bibnamefont {Zeng}}, \bibinfo {author} {\bibfnamefont
  {S.}~\bibnamefont {Hirai}}, \bibinfo {author} {\bibfnamefont {P.~P.}\
  \bibnamefont {Kong}}, \bibinfo {author} {\bibfnamefont {L.}~\bibnamefont
  {Li}}, \bibinfo {author} {\bibfnamefont {C.}~\bibnamefont {Park}}, \bibinfo
  {author} {\bibfnamefont {S.~J.}\ \bibnamefont {Yuan}}, \bibinfo {author}
  {\bibfnamefont {C.~Q.}\ \bibnamefont {Jin}}, \bibinfo {author} {\bibfnamefont
  {G.}~\bibnamefont {Cao}}, \ and\ \bibinfo {author} {\bibfnamefont {W.~L.}\
  \bibnamefont {Mao}},\ }\href
  {http://stacks.iop.org/0953-8984/26/i=21/a=215402} {\bibfield  {journal}
  {\bibinfo  {journal} {J. Phys.: Condens. Matter}\ }\textbf {\bibinfo {volume}
  {26}},\ \bibinfo {pages} {215402} (\bibinfo {year} {2014})}\BibitemShut
  {NoStop}%
\bibitem [{\citenamefont {Zocco}\ \emph {et~al.}(2014)\citenamefont {Zocco},
  \citenamefont {Hamlin}, \citenamefont {White}, \citenamefont {Kim},
  \citenamefont {Jeffries}, \citenamefont {Weir}, \citenamefont {Vohra},
  \citenamefont {Allen},\ and\ \citenamefont {Maple}}]{zocco2014persistent}%
  \BibitemOpen
  \bibfield  {author} {\bibinfo {author} {\bibfnamefont {D.~A.}\ \bibnamefont
  {Zocco}}, \bibinfo {author} {\bibfnamefont {J.~J.}\ \bibnamefont {Hamlin}},
  \bibinfo {author} {\bibfnamefont {B.~D.}\ \bibnamefont {White}}, \bibinfo
  {author} {\bibfnamefont {B.~J.}\ \bibnamefont {Kim}}, \bibinfo {author}
  {\bibfnamefont {J.~R.}\ \bibnamefont {Jeffries}}, \bibinfo {author}
  {\bibfnamefont {S.~T.}\ \bibnamefont {Weir}}, \bibinfo {author}
  {\bibfnamefont {Y.~K.}\ \bibnamefont {Vohra}}, \bibinfo {author}
  {\bibfnamefont {J.~W.}\ \bibnamefont {Allen}}, \ and\ \bibinfo {author}
  {\bibfnamefont {M.~B.}\ \bibnamefont {Maple}},\ }\href
  {http://stacks.iop.org/0953-8984/26/i=25/a=255603} {\bibfield  {journal}
  {\bibinfo  {journal} {J. Phys.: Condens. Matterr}\ }\textbf {\bibinfo
  {volume} {26}},\ \bibinfo {pages} {255603} (\bibinfo {year}
  {2014})}\BibitemShut {NoStop}%
\bibitem [{\citenamefont {Datchi}\ \emph {et~al.}(1997)\citenamefont {Datchi},
  \citenamefont {LeToullec},\ and\ \citenamefont
  {Loubeyre}}]{datchi1997improved}%
  \BibitemOpen
  \bibfield  {author} {\bibinfo {author} {\bibfnamefont {F.}~\bibnamefont
  {Datchi}}, \bibinfo {author} {\bibfnamefont {R.}~\bibnamefont {LeToullec}}, \
  and\ \bibinfo {author} {\bibfnamefont {P.}~\bibnamefont {Loubeyre}},\ }\href
  {\doibase http://dx.doi.org/10.1063/1.365025} {\bibfield  {journal} {\bibinfo
   {journal} {J. Appl. Phys.}\ }\textbf {\bibinfo {volume} {81}},\ \bibinfo
  {pages} {3333} (\bibinfo {year} {1997})}\BibitemShut {NoStop}%
\bibitem [{\citenamefont {Hammersley}\ \emph {et~al.}(1996)\citenamefont
  {Hammersley}, \citenamefont {Svensson}, \citenamefont {Hanfland},
  \citenamefont {Fitch},\ and\ \citenamefont {Hausermann}}]{Fit2D}%
  \BibitemOpen
  \bibfield  {author} {\bibinfo {author} {\bibfnamefont {A.~P.}\ \bibnamefont
  {Hammersley}}, \bibinfo {author} {\bibfnamefont {S.~O.}\ \bibnamefont
  {Svensson}}, \bibinfo {author} {\bibfnamefont {M.}~\bibnamefont {Hanfland}},
  \bibinfo {author} {\bibfnamefont {A.~N.}\ \bibnamefont {Fitch}}, \ and\
  \bibinfo {author} {\bibfnamefont {D.}~\bibnamefont {Hausermann}},\ }\href
  {\doibase 10.1080/08957959608201408} {\bibfield  {journal} {\bibinfo
  {journal} {High Press. Res.}\ }\textbf {\bibinfo {volume} {14}},\ \bibinfo
  {pages} {235} (\bibinfo {year} {1996})}\BibitemShut {NoStop}%
\bibitem [{\citenamefont {Rodriguez-Carvajal}(1993)}]{rodriguez1990fullprof}%
  \BibitemOpen
  \bibfield  {author} {\bibinfo {author} {\bibfnamefont {J.}~\bibnamefont
  {Rodriguez-Carvajal}},\ }\href {\doibase
  http://dx.doi.org/10.1016/0921-4526(93)90108-I} {\bibfield  {journal}
  {\bibinfo  {journal} {Physica B}\ }\textbf {\bibinfo {volume} {192}},\
  \bibinfo {pages} {55 } (\bibinfo {year} {1993})}\BibitemShut {NoStop}%
\bibitem [{\citenamefont {Momma}\ and\ \citenamefont {Izumi}(2008)}]{VESTA}%
  \BibitemOpen
  \bibfield  {author} {\bibinfo {author} {\bibfnamefont {K.}~\bibnamefont
  {Momma}}\ and\ \bibinfo {author} {\bibfnamefont {F.}~\bibnamefont {Izumi}},\
  }\href {\doibase 10.1107/S0021889808012016} {\bibfield  {journal} {\bibinfo
  {journal} {J. Appl. Crystallogr.}\ }\textbf {\bibinfo {volume} {41}},\
  \bibinfo {pages} {653} (\bibinfo {year} {2008})}\BibitemShut {NoStop}%
\bibitem [{\citenamefont {Matsuhata}\ \emph {et~al.}(2004)\citenamefont
  {Matsuhata}, \citenamefont {Nagai}, \citenamefont {Yoshida}, \citenamefont
  {Hara}, \citenamefont {ichi Ikeda},\ and\ \citenamefont
  {Shirakawa}}]{matsuhata2004crystal}%
  \BibitemOpen
  \bibfield  {author} {\bibinfo {author} {\bibfnamefont {H.}~\bibnamefont
  {Matsuhata}}, \bibinfo {author} {\bibfnamefont {I.}~\bibnamefont {Nagai}},
  \bibinfo {author} {\bibfnamefont {Y.}~\bibnamefont {Yoshida}}, \bibinfo
  {author} {\bibfnamefont {S.}~\bibnamefont {Hara}}, \bibinfo {author}
  {\bibfnamefont {S.}~\bibnamefont {ichi Ikeda}}, \ and\ \bibinfo {author}
  {\bibfnamefont {N.}~\bibnamefont {Shirakawa}},\ }\href {\doibase
  http://dx.doi.org/10.1016/j.jssc.2004.06.056} {\bibfield  {journal} {\bibinfo
   {journal} {J. Solid State Chem.}\ }\textbf {\bibinfo {volume} {177}},\
  \bibinfo {pages} {3776 } (\bibinfo {year} {2004})}\BibitemShut {NoStop}%
\bibitem [{\citenamefont {Singh}\ and\ \citenamefont
  {Nordstrom}(2006)}]{singh2006planewaves}%
  \BibitemOpen
  \bibfield  {author} {\bibinfo {author} {\bibfnamefont {D.~J.}\ \bibnamefont
  {Singh}}\ and\ \bibinfo {author} {\bibfnamefont {L.}~\bibnamefont
  {Nordstrom}},\ }\href@noop {} {\emph {\bibinfo {title} {Planewaves,
  Pseudopotentials, and the LAPW method}}}\ (\bibinfo  {publisher} {Springer
  Science \& Business Media},\ \bibinfo {year} {2006})\BibitemShut {NoStop}%
\bibitem [{\citenamefont {Giannozzi}\ \emph {et~al.}(2009)\citenamefont
  {Giannozzi}, \citenamefont {Baroni}, \citenamefont {Bonini}, \citenamefont
  {Calandra}, \citenamefont {Car}, \citenamefont {Cavazzoni}, \citenamefont
  {Ceresoli}, \citenamefont {Chiarotti}, \citenamefont {Cococcioni},
  \citenamefont {Dabo}, \citenamefont {Corso}, \citenamefont {de~Gironcoli},
  \citenamefont {Fabris}, \citenamefont {Fratesi}, \citenamefont {Gebauer},
  \citenamefont {Gerstmann}, \citenamefont {Gougoussis}, \citenamefont
  {Kokalj}, \citenamefont {Lazzeri}, \citenamefont {Martin-Samos},
  \citenamefont {Marzari}, \citenamefont {Mauri}, \citenamefont {Mazzarello},
  \citenamefont {Paolini}, \citenamefont {Pasquarello}, \citenamefont
  {Paulatto}, \citenamefont {Sbraccia}, \citenamefont {Scandolo}, \citenamefont
  {Sclauzero}, \citenamefont {Seitsonen}, \citenamefont {Smogunov},
  \citenamefont {Umari},\ and\ \citenamefont
  {Wentzcovitch}}]{giannozzi2009quantum}%
  \BibitemOpen
  \bibfield  {author} {\bibinfo {author} {\bibfnamefont {P.}~\bibnamefont
  {Giannozzi}}, \bibinfo {author} {\bibfnamefont {S.}~\bibnamefont {Baroni}},
  \bibinfo {author} {\bibfnamefont {N.}~\bibnamefont {Bonini}}, \bibinfo
  {author} {\bibfnamefont {M.}~\bibnamefont {Calandra}}, \bibinfo {author}
  {\bibfnamefont {R.}~\bibnamefont {Car}}, \bibinfo {author} {\bibfnamefont
  {C.}~\bibnamefont {Cavazzoni}}, \bibinfo {author} {\bibfnamefont
  {D.}~\bibnamefont {Ceresoli}}, \bibinfo {author} {\bibfnamefont {G.~L.}\
  \bibnamefont {Chiarotti}}, \bibinfo {author} {\bibfnamefont {M.}~\bibnamefont
  {Cococcioni}}, \bibinfo {author} {\bibfnamefont {I.}~\bibnamefont {Dabo}},
  \bibinfo {author} {\bibfnamefont {A.~D.}\ \bibnamefont {Corso}}, \bibinfo
  {author} {\bibfnamefont {S.}~\bibnamefont {de~Gironcoli}}, \bibinfo {author}
  {\bibfnamefont {S.}~\bibnamefont {Fabris}}, \bibinfo {author} {\bibfnamefont
  {G.}~\bibnamefont {Fratesi}}, \bibinfo {author} {\bibfnamefont
  {R.}~\bibnamefont {Gebauer}}, \bibinfo {author} {\bibfnamefont
  {U.}~\bibnamefont {Gerstmann}}, \bibinfo {author} {\bibfnamefont
  {C.}~\bibnamefont {Gougoussis}}, \bibinfo {author} {\bibfnamefont
  {A.}~\bibnamefont {Kokalj}}, \bibinfo {author} {\bibfnamefont
  {M.}~\bibnamefont {Lazzeri}}, \bibinfo {author} {\bibfnamefont
  {L.}~\bibnamefont {Martin-Samos}}, \bibinfo {author} {\bibfnamefont
  {N.}~\bibnamefont {Marzari}}, \bibinfo {author} {\bibfnamefont
  {F.}~\bibnamefont {Mauri}}, \bibinfo {author} {\bibfnamefont
  {R.}~\bibnamefont {Mazzarello}}, \bibinfo {author} {\bibfnamefont
  {S.}~\bibnamefont {Paolini}}, \bibinfo {author} {\bibfnamefont
  {A.}~\bibnamefont {Pasquarello}}, \bibinfo {author} {\bibfnamefont
  {L.}~\bibnamefont {Paulatto}}, \bibinfo {author} {\bibfnamefont
  {C.}~\bibnamefont {Sbraccia}}, \bibinfo {author} {\bibfnamefont
  {S.}~\bibnamefont {Scandolo}}, \bibinfo {author} {\bibfnamefont
  {G.}~\bibnamefont {Sclauzero}}, \bibinfo {author} {\bibfnamefont {A.~P.}\
  \bibnamefont {Seitsonen}}, \bibinfo {author} {\bibfnamefont {A.}~\bibnamefont
  {Smogunov}}, \bibinfo {author} {\bibfnamefont {P.}~\bibnamefont {Umari}}, \
  and\ \bibinfo {author} {\bibfnamefont {R.~M.}\ \bibnamefont {Wentzcovitch}},\
  }\href {http://stacks.iop.org/0953-8984/21/i=39/a=395502} {\bibfield
  {journal} {\bibinfo  {journal} {J. Phys.: Condens. Matter}\ }\textbf
  {\bibinfo {volume} {21}},\ \bibinfo {pages} {395502} (\bibinfo {year}
  {2009})}\BibitemShut {NoStop}%
\bibitem [{\citenamefont {Solovyev}\ \emph {et~al.}(1998)\citenamefont
  {Solovyev}, \citenamefont {Liechtenstein},\ and\ \citenamefont
  {Terakura}}]{solovyev1998hund}%
  \BibitemOpen
  \bibfield  {author} {\bibinfo {author} {\bibfnamefont {I.~V.}\ \bibnamefont
  {Solovyev}}, \bibinfo {author} {\bibfnamefont {A.~I.}\ \bibnamefont
  {Liechtenstein}}, \ and\ \bibinfo {author} {\bibfnamefont {K.}~\bibnamefont
  {Terakura}},\ }\href {\doibase 10.1103/PhysRevLett.80.5758} {\bibfield
  {journal} {\bibinfo  {journal} {Phys. Rev. Lett.}\ }\textbf {\bibinfo
  {volume} {80}},\ \bibinfo {pages} {5758} (\bibinfo {year}
  {1998})}\BibitemShut {NoStop}%
\bibitem [{\citenamefont {Subramanian}\ \emph {et~al.}(1994)\citenamefont
  {Subramanian}, \citenamefont {Crawford},\ and\ \citenamefont
  {Harlow}}]{subramanian1994single}%
  \BibitemOpen
  \bibfield  {author} {\bibinfo {author} {\bibfnamefont {M.}~\bibnamefont
  {Subramanian}}, \bibinfo {author} {\bibfnamefont {M.}~\bibnamefont
  {Crawford}}, \ and\ \bibinfo {author} {\bibfnamefont {R.}~\bibnamefont
  {Harlow}},\ }\href {\doibase http://dx.doi.org/10.1016/0025-5408(94)90120-1}
  {\bibfield  {journal} {\bibinfo  {journal} {Materials Research Bulletin}\
  }\textbf {\bibinfo {volume} {29}},\ \bibinfo {pages} {645 } (\bibinfo {year}
  {1994})}\BibitemShut {NoStop}%
\bibitem [{\citenamefont {Moretti~Sala}\ \emph
  {et~al.}(2014{\natexlab{a}})\citenamefont {Moretti~Sala}, \citenamefont
  {Rossi}, \citenamefont {Al-Zein}, \citenamefont {Boseggia}, \citenamefont
  {Hunter}, \citenamefont {Perry}, \citenamefont {Prabhakaran}, \citenamefont
  {Boothroyd}, \citenamefont {Brookes}, \citenamefont {McMorrow}, \citenamefont
  {Monaco},\ and\ \citenamefont {Krisch}}]{moretti2014crystal}%
  \BibitemOpen
  \bibfield  {author} {\bibinfo {author} {\bibfnamefont {M.}~\bibnamefont
  {Moretti~Sala}}, \bibinfo {author} {\bibfnamefont {M.}~\bibnamefont {Rossi}},
  \bibinfo {author} {\bibfnamefont {A.}~\bibnamefont {Al-Zein}}, \bibinfo
  {author} {\bibfnamefont {S.}~\bibnamefont {Boseggia}}, \bibinfo {author}
  {\bibfnamefont {E.~C.}\ \bibnamefont {Hunter}}, \bibinfo {author}
  {\bibfnamefont {R.~S.}\ \bibnamefont {Perry}}, \bibinfo {author}
  {\bibfnamefont {D.}~\bibnamefont {Prabhakaran}}, \bibinfo {author}
  {\bibfnamefont {A.~T.}\ \bibnamefont {Boothroyd}}, \bibinfo {author}
  {\bibfnamefont {N.~B.}\ \bibnamefont {Brookes}}, \bibinfo {author}
  {\bibfnamefont {D.~F.}\ \bibnamefont {McMorrow}}, \bibinfo {author}
  {\bibfnamefont {G.}~\bibnamefont {Monaco}}, \ and\ \bibinfo {author}
  {\bibfnamefont {M.}~\bibnamefont {Krisch}},\ }\href {\doibase
  10.1103/PhysRevB.90.085126} {\bibfield  {journal} {\bibinfo  {journal} {Phys.
  Rev. B}\ }\textbf {\bibinfo {volume} {90}},\ \bibinfo {pages} {085126}
  (\bibinfo {year} {2014}{\natexlab{a}})}\BibitemShut {NoStop}%
\bibitem [{\citenamefont {Moretti~Sala}\ \emph
  {et~al.}(2014{\natexlab{b}})\citenamefont {Moretti~Sala}, \citenamefont
  {Rossi}, \citenamefont {Boseggia}, \citenamefont {Akimitsu}, \citenamefont
  {Brookes}, \citenamefont {Isobe}, \citenamefont {Minola}, \citenamefont
  {Okabe}, \citenamefont {R\o{}nnow}, \citenamefont {Simonelli}, \citenamefont
  {McMorrow},\ and\ \citenamefont {Monaco}}]{moretti2014orbital}%
  \BibitemOpen
  \bibfield  {author} {\bibinfo {author} {\bibfnamefont {M.}~\bibnamefont
  {Moretti~Sala}}, \bibinfo {author} {\bibfnamefont {M.}~\bibnamefont {Rossi}},
  \bibinfo {author} {\bibfnamefont {S.}~\bibnamefont {Boseggia}}, \bibinfo
  {author} {\bibfnamefont {J.}~\bibnamefont {Akimitsu}}, \bibinfo {author}
  {\bibfnamefont {N.~B.}\ \bibnamefont {Brookes}}, \bibinfo {author}
  {\bibfnamefont {M.}~\bibnamefont {Isobe}}, \bibinfo {author} {\bibfnamefont
  {M.}~\bibnamefont {Minola}}, \bibinfo {author} {\bibfnamefont
  {H.}~\bibnamefont {Okabe}}, \bibinfo {author} {\bibfnamefont {H.~M.}\
  \bibnamefont {R\o{}nnow}}, \bibinfo {author} {\bibfnamefont {L.}~\bibnamefont
  {Simonelli}}, \bibinfo {author} {\bibfnamefont {D.~F.}\ \bibnamefont
  {McMorrow}}, \ and\ \bibinfo {author} {\bibfnamefont {G.}~\bibnamefont
  {Monaco}},\ }\href {\doibase 10.1103/PhysRevB.89.121101} {\bibfield
  {journal} {\bibinfo  {journal} {Phys. Rev. B}\ }\textbf {\bibinfo {volume}
  {89}},\ \bibinfo {pages} {121101} (\bibinfo {year}
  {2014}{\natexlab{b}})}\BibitemShut {NoStop}%
\bibitem [{\citenamefont {Moser}\ \emph {et~al.}(2014)\citenamefont {Moser},
  \citenamefont {Moreschini}, \citenamefont {Ebrahimi}, \citenamefont
  {Dalla~Piazza}, \citenamefont {Isobe}, \citenamefont {Okabe}, \citenamefont
  {Akimitsu}, \citenamefont {Mazurenko}, \citenamefont {Kim}, \citenamefont
  {Bostwick}, \citenamefont {Rotenberg}, \citenamefont {Chang}, \citenamefont
  {R{\o}nnow},\ and\ \citenamefont {Grioni}}]{moser2014electronic}%
  \BibitemOpen
  \bibfield  {author} {\bibinfo {author} {\bibfnamefont {S.}~\bibnamefont
  {Moser}}, \bibinfo {author} {\bibfnamefont {L.}~\bibnamefont {Moreschini}},
  \bibinfo {author} {\bibfnamefont {A.}~\bibnamefont {Ebrahimi}}, \bibinfo
  {author} {\bibfnamefont {B.}~\bibnamefont {Dalla~Piazza}}, \bibinfo {author}
  {\bibfnamefont {M.}~\bibnamefont {Isobe}}, \bibinfo {author} {\bibfnamefont
  {H.}~\bibnamefont {Okabe}}, \bibinfo {author} {\bibfnamefont
  {J.}~\bibnamefont {Akimitsu}}, \bibinfo {author} {\bibfnamefont {V.~V.}\
  \bibnamefont {Mazurenko}}, \bibinfo {author} {\bibfnamefont {K.~S.}\
  \bibnamefont {Kim}}, \bibinfo {author} {\bibfnamefont {A.}~\bibnamefont
  {Bostwick}}, \bibinfo {author} {\bibfnamefont {E.}~\bibnamefont {Rotenberg}},
  \bibinfo {author} {\bibfnamefont {J.}~\bibnamefont {Chang}}, \bibinfo
  {author} {\bibfnamefont {H.~M.}\ \bibnamefont {R{\o}nnow}}, \ and\ \bibinfo
  {author} {\bibfnamefont {M.}~\bibnamefont {Grioni}},\ }\href@noop {}
  {\bibfield  {journal} {\bibinfo  {journal} {New J. Phys.}\ }\textbf {\bibinfo
  {volume} {16}},\ \bibinfo {pages} {013008} (\bibinfo {year}
  {2014})}\BibitemShut {NoStop}%
\bibitem [{\citenamefont {Solovyev}\ \emph {et~al.}(2015)\citenamefont
  {Solovyev}, \citenamefont {Mazurenko},\ and\ \citenamefont
  {Katanin}}]{solovyev2015validity}%
  \BibitemOpen
  \bibfield  {author} {\bibinfo {author} {\bibfnamefont {I.~V.}\ \bibnamefont
  {Solovyev}}, \bibinfo {author} {\bibfnamefont {V.~V.}\ \bibnamefont
  {Mazurenko}}, \ and\ \bibinfo {author} {\bibfnamefont {A.~A.}\ \bibnamefont
  {Katanin}},\ }\href {\doibase 10.1103/PhysRevB.92.235109} {\bibfield
  {journal} {\bibinfo  {journal} {Phys. Rev. B}\ }\textbf {\bibinfo {volume}
  {92}},\ \bibinfo {pages} {235109} (\bibinfo {year} {2015})}\BibitemShut
  {NoStop}%
\end{thebibliography}
\end{document}